\documentclass{article}
\usepackage[final]{neurips_2024}

\usepackage{pdfpages}
\usepackage[utf8]{inputenc} 
\usepackage[T1]{fontenc}    
\usepackage{hyperref}       
\usepackage{url}            
\usepackage{booktabs}       
\usepackage{amsfonts}       
\usepackage{nicefrac}       
\usepackage{microtype}      
\usepackage{xcolor}
\usepackage{natbib}
\usepackage{amssymb}
\usepackage{relsize}
\usepackage{amsmath}
\usepackage{bbding}

\definecolor{Green}{rgb}{0, 0.6, 0.1}
\definecolor{Orange}{rgb}{1, 0.6, 0}

\usepackage{tikz}
\newcommand{\tikzxmark}{%
\tikz[scale=0.23] {
    \draw[line width=1.1,line cap=round] (0,0) to [bend left=6] (1,1);
    \draw[line width=1.1,line cap=round] (0.2,0.95) to [bend right=3] (0.8,0.05);
}}

\title{A Turing Test for Artificial Nets\\ devoted to model Human Vision}

\author{Jorge Vila-Tomás\\
  Image Processing Lab\\
  Universitat de València\\
  Paterna 46980, València, Spain \\
  \texttt{jorge.vila-tomas@uv.es} \\
  \And
  Pablo Hernandez-Cámara\\
  Image Processing Lab\\
  Universitat de València\\
  \texttt{pablo.hernandez-camara@uv.es} \\
  \And
  Qiang Li\\
  Image Processing Lab\\
  Universitat de València\\
  TReNDS\\
  Georgia State, Georgia Tech, and Emory\\ 
  \texttt{qiang.li@uv.es} \\
  \And
  Valero Laparra\\
  Image Processing Lab\\
  Universitat de València\\
  \texttt{valero.laparra@uv.es}\\  
  \And
  Jesús Malo\thanks{Corresponding author. Web Site: \texttt{https://isp.uv.es/excathedra.html}
  } \\
  Image Processing Lab\\
  Universitat de València\\
  \texttt{jesus.malo@uv.es} \\  
  }

\begin{document}
\maketitle

\vspace{-0.4cm}
\begin{abstract}
  In this work\footnote{Concept and results first presented at the \emph{AI Evaluation Workshop} at the University of Bristol, June 2022.}
  we argue that, despite recent claims about successful modeling of the visual brain using deep nets, the problem is far from being solved, particularly for low-level vision. 
  Open issues include \emph{where should we read from in ANNs to check behavior?}, \emph{what should be the read-out?},  \emph{this ad-hoc read-out is considered part of the brain model or not?}, in order to understand vision-ANNs, \emph{should we use artificial psychophysics or artificial physiology?}, anyhow, \emph{artificial tests should literally match the experiments done with humans?}. These questions suggest a clear need of biologically sensible tests for deep models of the visual brain, and more generally, to understand ANNs devoted to generic vision tasks. 
  
  Following our use of low-level facts from \emph{Vision Science} in image processing, we present a low-level dataset compiling the basic spatio-temporal and chromatic facts that describe the adaptive information bottleneck of the retina-V1 pathway, and are not currently available in popular databases such as BrainScore. We propose its use for model evaluation.
  
  As illustration of the proposed methods we check the behavior of three recent models with similar deep architecture: \textbf{(1)}~A parametric model tuned via the psychophysical method of Maximum Differentiation [Malo \& Simoncelli SPIE 15, Martinez et al. PLOS 18, Martinez et al. Front. Neurosci. 19], \textbf{(2)}~A non-parametric model (the \emph{PerceptNet}) tuned to maximize the correlation with humans on subjective image distortions [Hepburn et al. IEEE ICIP 20], and \textbf{(3)}~A~model with the same encoder as the \emph{PerceptNet}, but tuned for image segmentation [Hernandez-Camara et al. Patt.Recogn.Lett. 23, Hernandez-Camara et al. Neurocomp. 25]. Results on 10 compelling psycho/physio visual facts show that the first (parametric) model is the one with closer behavior to humans in  terms of the nonlinear behavior when facing complex spatio-chromatic patterns.  
\end{abstract}

\clearpage
\section{Introduction}

\vspace{-0.05cm}
\subsection{Prologue}
This work reproduces our \emph{talk} (otherwise unpublished in print) at the AI Evaluation Workshop in june 2022 at the AI Dept. of the University of Bristol organized by Prof. Raul Santos of the Eng. Maths Dept. of UoB.
That \emph{talk}, that proposed an original methodology (with experimental results) to evaluate deep nets devoted to vision tasks, was the seed of our current work with Prof. Jeff Bowers of the Psychol. Dept. of UoB in the context of the Benjamin Meaker Distinguished Professorship granted to Prof. Jesús Malo in 2024, as a low-level complement to the (high-level) Bowers' proposals in~\cite{bowers2023deep,Bowers24}. 
Journal publication of this 2022 \emph{talk}, is pertinent for a wider audience because this approach based in low-level visual psychophysics is still unusual in the AI and machine learning communities, despite some researchers are independently proposing very similar evaluations quite recently~\cite{Rafal_CVPR_25,Rafal25}.
As shown below, our proposed evaluation program includes facts that go beyond the luminance, color, and contrast masking  facts considered in~\cite{Rafal_CVPR_25,Rafal25}. The work of Rafal Mantiuk's lab shares the same spirit and focus on low-level psychophysics, but his \emph{more quantitative comparison} is in contrast with our proposal, which stress the \emph{qualitative understanding} of the human response curves so that the AI researchers can spot major conceptual errors in deep models in an easy way. Moreover, as explained below, the selected visual stimuli\footnote{All made online available here since 2022: \texttt{http://isp.uv.es/docs/TuringTestVision.zip}} (and associated psychophysical facts\footnote{Original 2022 slides (UoB AI Evaluation Workshop): \texttt{http://isp.uv.es/docs/talk\_AI\_Bristol\_Malo\_et\_al\_2022.pdf}}), allow to intuitively infer modifications in the architectures in order to correct the detected errors.

\vspace{-0.05cm}
\subsection{Motivation: is that model really human-like?}
The motivation for our proposal starts by reviewing the claims about how deep learning models are the ultimate tool to model the visual brain, as recalled in~\cite{bowers2023deep}. Claims cited by Bowers et al. include~\cite{ICLR2019,PNAS2021,otroPNAS2021,JournalCognSci2021,JNeurosci18,NeuralNets2021}. Other examples in the same vein not cited by Bowers et al. include~\cite{PLOSCompBiol2019,PLOSCompBiol2020}.
Nevertheless, following the skeptical tone of Bowers et al.~\cite{bowers2023deep}, the crucial comment made by some scientists (for example at the Center of Neural Science of NYU after they carefully listen to the details of your model) is \emph{yes, yes, that is nice, but the brain doesn't work like that, does it?}~\cite{PaninskiPersonal01}.

Two additional examples of the skepticism in that critical question include \emph{Tomaso Poggio} and \emph{Horace Barlow}. 
In the 70's David Marr and Tomaso Poggio proposed an interesting taxonomy of the approaches to the vision problem: their famous \emph{separate abstraction levels}, namely, computational, algorithmic and implementation~\cite{MarrPoggio77,Marr78}. 
However, 42 years later, in view of the current tools to optimize models, Poggio himself questioned the separability of these levels~\cite{Poggio21}. This taxonomy has been inspiring for decades, but now it is under dabate~\cite{Lengyel24,Pillow24,Malo24a}. For example, work on color illusions~\cite{GomezVilla20} and CSFs in autoencoders~\cite{Li22}, and work on subjective distances between images in ANNs~\cite{Hernandez25,Hepburn22} stress the relation between the computational and the algorithmic levels, thus questioning previous (purely computational) explanations  that disregard architecture~\cite{Malo06b,Laparra12,Laparra15}. 
In a similar vein, Horace Barlow, 50 years after his inspiring \emph{Efficient Coding Hypothesis}~\cite{Barlow59,Barlow61}, questioned purely infomax approaches~\cite{Barlow01}, and, for a similar reason, he questioned our preliminary work 
on the use of Principal Curves to explain color and texture nonlinearities of human vision based on the data, back in 2004~\cite{GRCmeetingOxford04}: \emph{that is interesting, but the visual brain may not work like that}~\cite{BarlowPersonal04}.

That skepticism (based on the range of empirical behavior explained and the assumptions made to make these explanations) is the core of the spirit in~\cite{bowers2023deep}, and also the motivation of this work. 
Therefore, the key ideas of this work are basically two:

\vspace{-0.3cm}
\begin{itemize}

\item The use of AI techniques (e.g. deep learning) to understand the visual brain may not be as easy as people thought back in 2022, and even now. More explanatory tests are required.

\item Our specific proposal here is a low-level Turing test based on 10-points low-level physiological and psychophysical facts (our \emph{Decalogue}) to check if certain artificial model behaves as the (low-level) human visual brain.

\end{itemize}

\subsection{Structure of the paper}

Section~2 states that the question \emph{are the models sensible from the point of view of low-level physiology and psychophysics?} remains open from the perspective of modeling and evaluation.
In Section~3, we propose our contribution: an easy-to-use test (consisting on online available visual stimuli) and associated responses illustrated here for evaluation of deep learning vision models. 
These stimuli visually illustrate low-level phenomena described by classical \emph{Vision Science}.
In Section~4, we illustrate the proposed method through the original evaluation of three recent models: 
(1)~a classically formulated, not end-to-end optimized, model with functional form derived from classical vision science literature, where the specific values of its parameters have been psychophysically measured~\cite{Malo15,Martinez18,Martinez19,Malo24}.
(2)~A network with a bio-inspired architecture but with free parameters end-to-end optimized to reproduce subjective image quality, the \emph{PerceptNet}~\cite{Hepburn20}.
It resembles AlexNet and VGG, but it was specifically designed to accommodate the known aspects of the retina-cortex visual pathway using a constrained version of divisive normalization~\cite{Balle17}.
And, (3) a model with the same style encoder as the \emph{PerceptNet}, but augmented with a decoder, and both (encoder and decoder) are trained for image segmentation~\cite{Hernandez23,Hernandez24}, which is also a biologically plausible task.
Section~5 discusses what can be generally obtained from the proposed test methods (which might be considered qualitative). Note that 
even in the engineering case where one does not necessarily need the networks to resemble human vision, one would always want them to have good adaptation properties to achieve good generalization, and potential failures in this regard become clearly evident through the proposed tests.
Finally, Section~6 concludes the paper. 


\section{Open issues in modeling vision}

As pointed out in~\cite{Torralba24} the basic question, as in human vision, is how to deal with deep models which are hardly explainable black-boxes once trained. 

\subsection{Uncertain computational goal}

\begin{figure}[b]
\begin{center}
\includegraphics[width=0.8\textwidth]{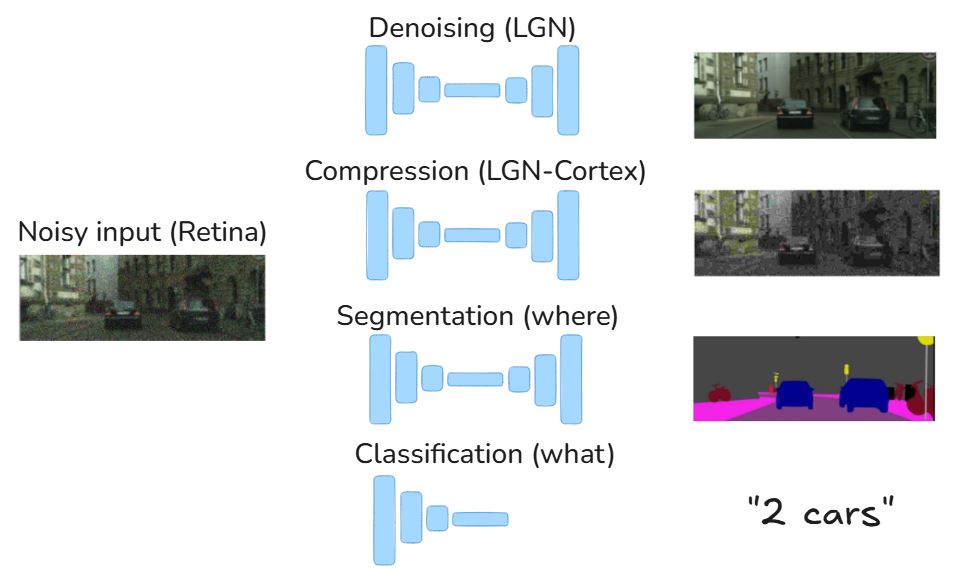}
\vspace{-0.3cm}
\caption{\small{Image Denoising, Image Compression, Image Segmentation and Image Classification architectures with (eventually) biological correlates in the LGN, the V1 and beyond. However, it is not obvious how how these tasks may be combined to explain biological vision.}}
\label{ejemplos1}
\vspace{-0.3cm}
\end{center}
\end{figure}

First, the more general open issue is the discussion on the \emph{computational goal} that eventually explains the organization and behavior of visual systems. Consider architectures/tasks such as the ones presented in Fig.~\ref{ejemplos1}. These tasks are related to low-, mid-, and high-level tasks arguably implemented by biological vision.
In biology enhancement of the blurry and noisy signal in the retina has been proposed as an explanation of the the LGN, as pursuing this goal may reproduce some of its spatio-chromatic~\cite{Li92,Li22} and purely chromatic~\cite{GomezVilla20} features. Another example is the compression possibly happening in part at the LGN bottleneck and at the feature selection after V1.
Bandwidth limitation, dimensionality reduction and attention focus are sensible goals in this regard~\cite{Karklin11,lindsey19,Li14}. 
A number of compression algorithms (for images~\cite{Wallace92,Malo95,Malo00b,Marcellin01,Malo06a,Balle17} and video~\cite{Legall92,Malo00a,Malo00c,Malo01a}) have been based on human vision models. 
Segmentation is arguably another (mid-level) task that has to be done by biological vision, and biological nonlinearities have been shown to improve segmentation in images~\cite{Hernandez23,Hernandez24} and video~\cite{Malo00c,Malo01a}. Arguably, segmentation is implemented in the \emph{where} channel from the lower-level primitives extracted in V1~\cite{Goodale91,Milner92}. Higher-level tasks such as classification are supposed to happen in the \emph{what} channel~\cite{Logothetis96,Koch00}. And similarly, in standard models such as the one depicted in Fig.~\ref{ejemplos1} biological nonlinearities have been shown to have a significant role in classification~\cite{Coen13,Miller22}.

\subsection{Uncertain read-out mechanisms}

\begin{figure}[b]
\begin{center}
\vspace{-0.75cm}
\includegraphics[width=0.5\textwidth]{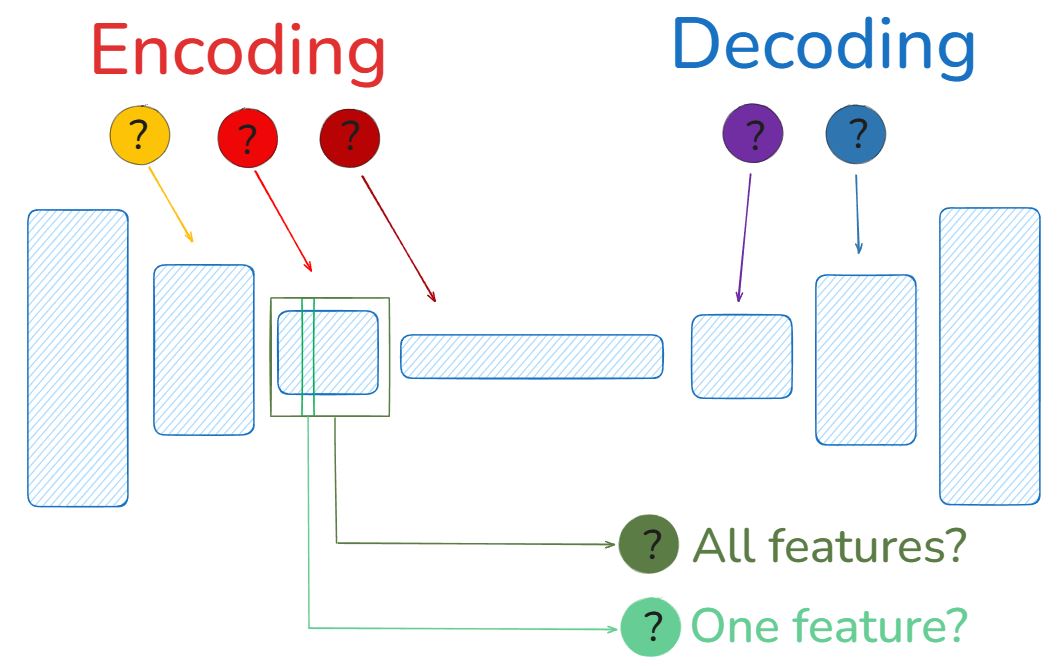}
\vspace{-0.3cm}
\caption{\small{Given a deep model successfully trained for some visual task, the read-out location and read-out mechanism (or decoder) is important to assess its biological plausibility.}}
\label{where-and-How}
\vspace{-0.5cm}
\end{center}
\end{figure}

As stated in the introduction, in the age of automatic differentiation where the classical Marr-Poggio levels are not that separated, the \emph{computational goal} is not the only open issue. 
For instance, in order to check if a (mathematical) model is biologically sensible, where should we read the signals from? The read-out mechanism is also important.
Note that the fact that certain layer has the necessary information in order to solve a task (read-out in \emph{any complicated} way, e.g. a highly specialized dense network) is not enough to say that this layer represents the way the visual brain works: the necessary information is already present in the retina (if read in proper way) and, of course, the retina is not a good model for the rest of the visual brain. This problem is illustrated in Fig.~\ref{where-and-How}.

\begin{figure}[t]
\begin{center}
\vspace{-0.75cm}
\includegraphics[width=0.95\textwidth]{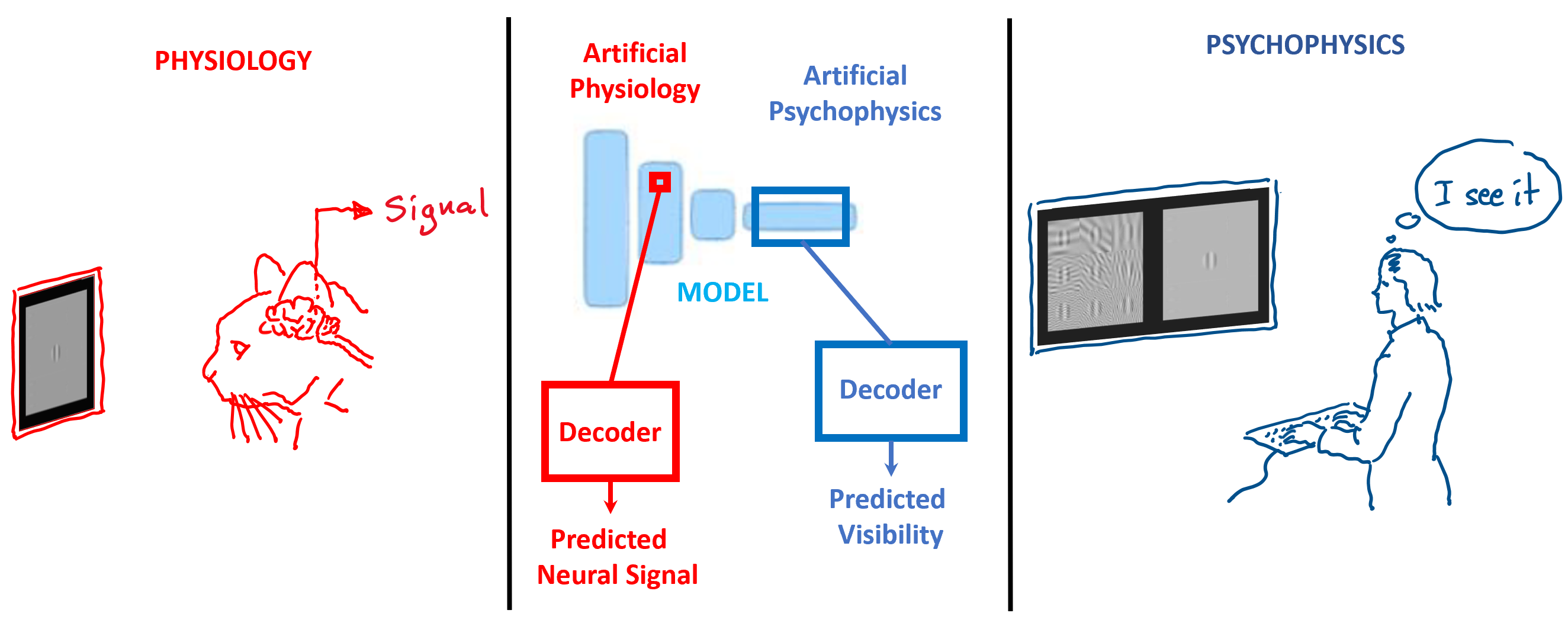}
\vspace{-0.3cm}
\caption{\small{In artificial physiology (left) and in artificial psychophysics (right), the arbitrary decoder to read-out model activations is critical.}}
\label{physio_psycho_read}
\vspace{-0.0cm}
\end{center}
\end{figure}

In the case of doing \emph{artificial physiology}, i.e. reading the signals from certain neurons or layer, or \emph{artificial psychophysics}, i.e. trying to make decisions from the responses of the network, for instance to decide if certain stimulus is visible or not, one should propose a \emph{read-out mechanism} to summarize the responses into a decision variable (see Fig.~\ref{physio_psycho_read}). The selection of the \emph{read-out mechanism} is not trivial.
In fact, the quality of the read-out information may strongly depend on the complexity of this (arbitrarily selected) mechanism. As a result, one may not be able to tell if the model itself is good, or the good behavior has to be attributed to a clever read-out which is not part of the model.
Examples include the use of classifiers at certain location of the network to make a decision on visibility, as in~\cite{Coen13,Akbarinia23}, or without clasifiers relying on the model output\cite{CCN25}; or the (more classical) use of Euclidean distances between stimuli to tell if they are discriminable~\cite{Teo94a,Li22,Hernandez25}. These (arbitrary) decisions definitely affect the characterization of the system, e.g. its frequency response~\cite{Li22,Akbarinia23}. For example, linear or nonlinear classifiers effectively apply different (non-Euclidean) distance metrics~\cite{Duda73} and, hence, they should lead  to different decisions. 

Another (more particular) discussion is the debate on the summation, which is classical in vision science~\cite{graham_visual_1989}: for instance, which Minkowski exponent is more physiologically plausible?. Note that using different norms and summation schemes definitely lead to different results~\cite{Laparra10}. 
A final (also non obvious) way of assessing stimuli in the network is measuring differences in the statistical properties of the response~\cite{Wang03,Ding20} or measuring information flow along the network~\cite{Sheikh05,Sheikh06,Malo20,Malo21,Li24}. These options require making non-trivial decisions such as which statistical descriptors make sense~\cite{Gatys16,Ding20}, or how to set the level of noise in the network~\cite{Sheikh05,Sheikh06,Malo20}. In this regard, models can be improved either by changing the architecture and the measures of information~\cite{Malo21,Laparra24}, or by better estimations of the internal noise~\cite{Malo25}. 

\subsection{Uncertain experimental setting}

\begin{figure}[b]
\begin{center}
\vspace{-0.0cm}
\includegraphics[width=1.0\textwidth]{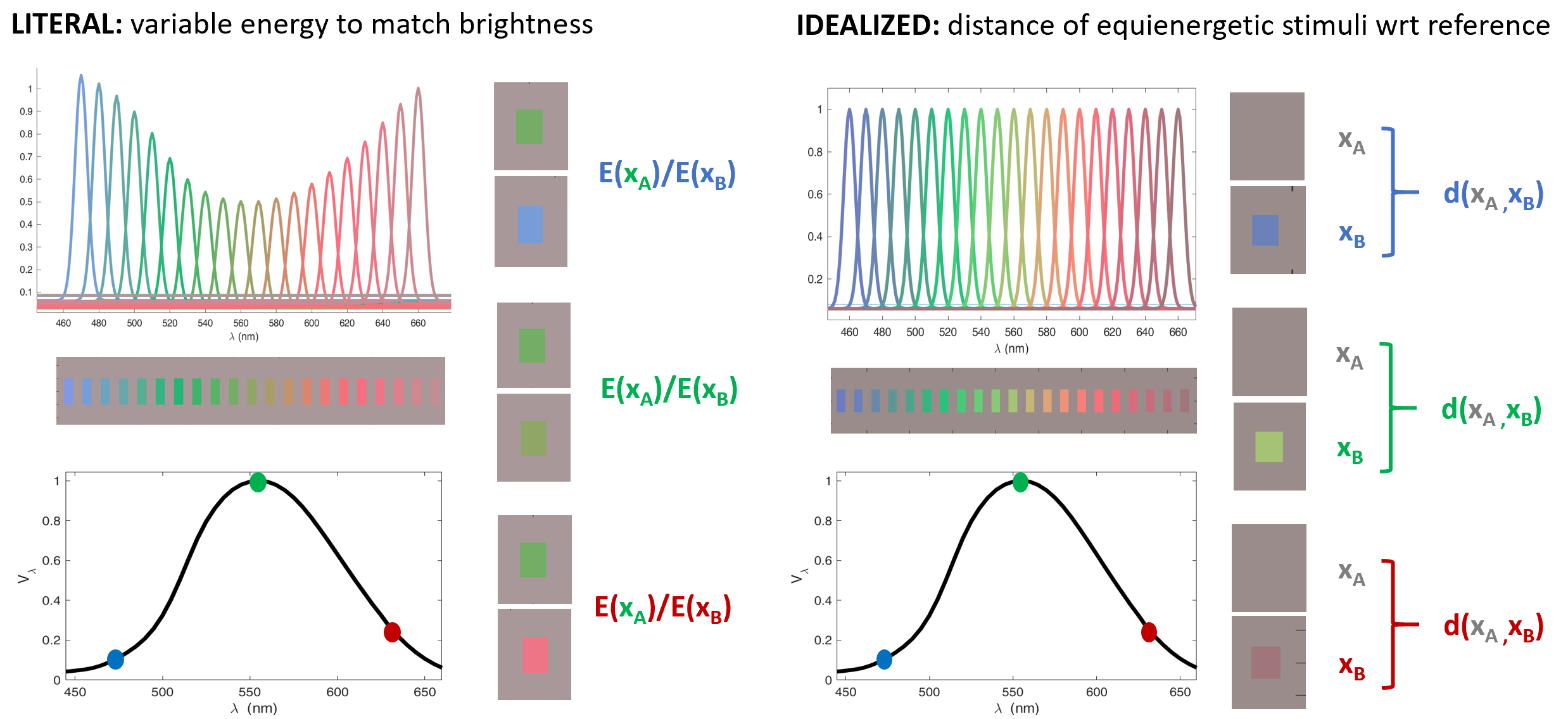}
\vspace{-0.3cm}
\caption{\small{In measuring the spectral sensitivity of certain elements of a network one may try a \emph{literal} reproduction of the human psychophysics (left) or an \emph{idealized} experiment (right). 
The \emph{literal reproduction} could be done through matching experiments~\cite{Stiles00}: finding the ratio of energies necessary to match the response to quasi-spectral stimuli of different wavelengths. 
The \emph{idealized} version of the experiment could be based on measuring the increment of response (distance) due to equal energy quasi-monochromatic stimuli with regard to a common reference.}}
\label{spectral}
\vspace{-0.5cm}
\end{center}
\end{figure}

And finally, the third open issue is the way of doing the evaluation: \emph{the experiment implementation matters}.
In particular, \emph{should we use artificial physiology or artificial psychophysics?}. 
Current techniques by the machine learning community to visualize the behavior of the networks~\cite{Vedaldi16,Luo16} are based on classical single cell recordings, such as the very concept of \emph{receptive field}~\cite{Hubel59,Hubel61,Ringach02}, and the identification of sensitive neurons by looking to the stimulus that maximizes the neuron response, which is a common practice in visual neuroscience~\cite{Lennie08}. However, there are more sophisticated techniques such as \emph{reverse correlation} which are used both in physiology~\cite{Ringach04} and in psychophysics~\cite{Eckstein02}, and these are not yet widely used in machine learning. 
Regarding the experimental setting, \emph{should one go for a literal reproduction of the experiments with humans, or should one try an idealized version of the experiment?}. This open question can be illustrated by the example in Fig.~\ref{spectral} on the  spectral sensitivity of a network.

This is a non-trivial question because, for instance, some techniques to assess visual illusions in a model involve the inversion of the inner representation~\cite{Otazu10,GomezVilla20}, 
which \emph{does not happen in the human brain} while others, similarly to human psychophysics~\cite{WareCowan82}, are based on \emph{matching} the response at the inner representation~\cite{gomez-villa_color_2020,GomezVilla25}. As stated above, this has implications as deciding at which layer one should impose the matching (or where to read from). 

\subsection{Better evaluation techniques are needed}

All these non-trivial decisions (despite they all belong to low-level characterizations of the visual system) clearly point out the need of better methodologies for model evaluation in order to assess how close different models may be to the visual brain.

In this context, our proposal here is simple: \emph{just provide the code to generate a set of well selected stimuli that illustrates a number of classical low-level visual psychophysics facts and have them prepared as inputs to evaluate image-computable models}.
The first version of such \emph{low-level Turing Test} (back in 2022) included stimuli for 10 well known behaviors (our \emph{Decalogue}). That decalogue is being extended to 20 facts in our (by 2025) on-going collaboration with Prof. Bowers~\cite{Malo24charla}. 

The selected stimuli (which include color, texture, and motion) are behind the current understanding of early vision as a set of linear-nonlinear layers~\cite{Rust06,Schutt17,Martinez18,Martinez19,Bertalmio20,Malo24,Bertalmio24}.
Our proposal follows the tradition of previous (too simple) low-level datasets such as the OSA ModelFest initiative~\cite{Modelfest}, but, low-level pychophysics has not been extensively included in the (today popular) BrainScore~\cite{BrainScore}, nor in the high-level criticisms done by Bowers et al.~\cite{bowers2023deep,Bowers24}.

\section{Our Proposal: A low-level vision Turing test for deep-nets}

\subsection{The Decalogue: facts and foundations}

The set of facts and associated stimuli included in our proposal is summarized Table~\ref{tabla1}. Among the rich literature of low-level visual psychophysics, the selection of those specific facts (and associated stimuli) is founded in two main reasons.

\textbf{First}, they describe the visual information adaptively captured (and discarded) by the front-end of human vision. 
On the one hand, linear sensitivities describe the spectral, chromatic, and spatio-temporal bandwidth and relative weight given by the system to the frequency components of the input stimuli. This linear description in terms of sensitivity filters is the first order approximation to the visual bottleneck. More interestingly, this bottleneck is adaptive: then in classical models of vision science, extra nonlinear mechanisms are proposed to be in between the linear filters to account for the adaptive responses to the specific eigen-stimuli of the linear filters. The stimuli in the test we compile here were specifically designed to probe those linear and nonlinear mechanisms of human vision.
The power and relevance of the selected stimuli for a complete characterization of the low-level bottleneck of image-computable models is suggested by the fact that, for decades, the straightforward use of these facts (with minor or no optimization at all) led to competitive image~\cite{Wallace92,Malo95,Malo00b,Marcellin01,Malo06a} and video coding algorithms~\cite{Legall92,Malo00a,Malo00c,Malo01a} and distortion metrics~\cite{Daly93,Watson93libro,Teo94a,Malo97,Watson02,Laparra10} equipped with color constancy and contrast 
adaptation~\cite{Hernandez23,Hernandez24,Hernandez24front}. Checking if the response of a network is human-like for those stimuli would imply that the bottleneck of the network would have \emph{statistically} good adaptive behavior~\cite{Olshausen96,Schwartz01,Malo06b,Malo10,Laparra12,Laparra15,Gomez19,Malo20,Malo22}.

\textbf{Second}, effects elicited by the selected stimuli are visually compelling and hence, the user of the test can check (by the eye) if the model under consideration behaves like humans or not. On the one hand, sensitivity surfaces to simple (isolated) stimuli are standardized and ready for direct quantitative comparison~\cite{Stiles00,Jameson57,Campbell68,Mullen85,Georgeson75,Daly90,Kelly79,Diez11}. On the other hand, as illustrated below, nonlinear responses when using stimuli in a context (under adaptation) have specific qualitative behaviors that are easy to see. In this way, that eventual model deviations from human-like behavior are easy to detect.

\begin{table}[b]
\begin{center}
\hspace{-2cm}\begin{tabular}{ccccc}
  & Facts      & Stimuli & Modality & Response  \\\hline
1 & Spectral Sensitivities (achromatic and opponent) & Quasi-spectral & Color & Linear \\
2 & Brightness \& Color Response Saturation & Color calibrated & Color & Non-linear \\
3 & Achromatic Contrast Sensitivity (Bandwidth) & Achrom. Gabors/noise & Texture & Linear \\
4 & Chromatic Contrast Sensitivity (Bandwidth) & Chrom. Gabors/noise & Texture & Linear \\
5 & Spatio-Chromatic Receptive Fields & Deltas / noise & Texture & Linear \\
6 & Nonlinear Contrast Response: Saturation & Gabors/noise & Texture & Non-linear \\
7 & Nonlinear Contrast Response: Frequency order & Gabors/noise & Texture & Non-linear \\
8 & Context effects: Energy & Gabors/noise & Texture & Non-linear \\
9 & Context effects: Frequency & Gabors/noise & Texture & Non-linear \\
10 & Context effects: Orientation & Gabors/noise & Texture & Non-linear \\
\end{tabular}
\vspace{0.3cm}
\caption{\small{Facts (and associated stimuli) of our Decalogue that are behind the current understanding of the information bottleneck happening between the retina and the V1 cortex. They include the color, texture and motion processing abilities of human early vision. In the original literature the stimuli were specifically designed to probe the linear or the nonlinear behavior of the system.}}
\label{tabla1}
\end{center}
\end{table}


\subsection{The Decalogue: specific examples}

In this section we show four examples of the proposed Decalogue with series of calibrated stimuli (from the colorimetric and the spatial perspectives) that illustrate the nonlinear response of humans to (i) luminance in different backgrounds leading to different perceptions of \emph{brightness}, (ii) deviations in opponent color directions under different induction conditions leading to different perceptions of \emph{hue} and \emph{saturation}, (iii) texture masking due to the energy of the background, and (iv) texture masking due to the similarity between the features of the background and test.   

It is important to note that the facts illustrated here (facts 2, 8, 10) are examples of the curves that are not standardized, as opposed to other 
facts in the proposed Decalogue (facts 1, 3, 4, 6, 7), in which strict quantitative comparisons are possible. The fact that, even in these non-standardized examples, the qualitative behavior is so compelling implies that the associated stimuli are useful to check, rank, and eventually rule out, artificial models.

\subsubsection{Luminance and brightness} 

The first set of stimuli refers to series of luminance-calibrated achromatic samples that illustrate the perception of brightness in backgrounds of different luminance.
They illustrate the Weber law~\cite{Stiles00,Fairchild13} and the crispening effect~\cite{Whittle92}, i.e. the achromatic part of fact~2 in Table~\ref{tabla1}.
These effects have been related to the statistics of natural images~\cite{Laughlin81,Laparra12} and with sophisticated models of retinal adaptation~\cite{Bertalmio20}.

\begin{figure}[b]
\begin{center}
\vspace{-0.75cm}
\includegraphics[width=0.95\textwidth]{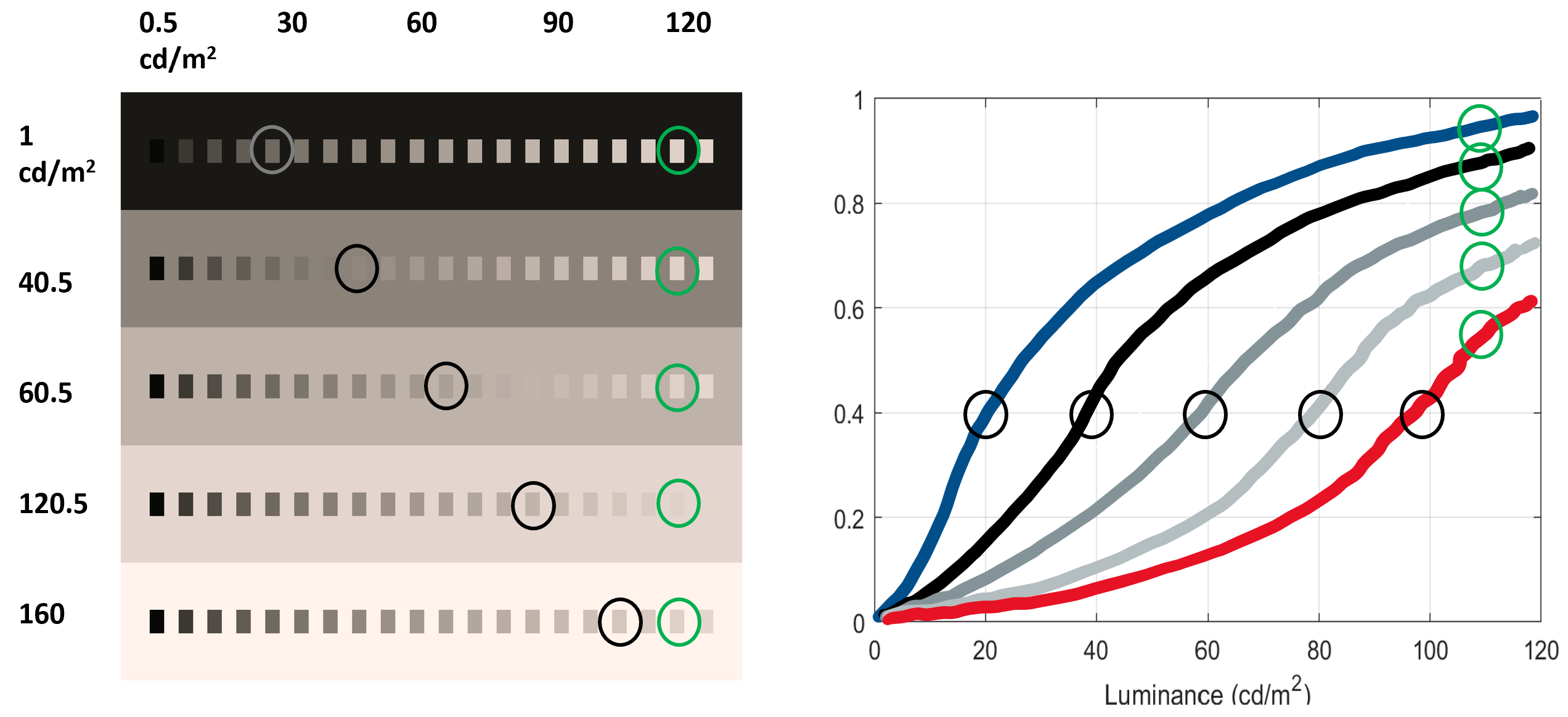}
\vspace{-0.3cm}
\caption{\small{Series of stimuli eliciting nonlinear brightness perception. Luminance calibrated linearly spaced tests in different luminance calibrated backgrounds. This illustrates the Weber  law~\cite{Fairchild13} as well as Whittle's crispening effect~\cite{Whittle92}, as summarized in~\cite{Bertalmio20}.}}
\label{weber}
\vspace{-0.9cm}
\end{center}
\end{figure}

Figure~\ref{weber} show a series of these stimuli in the (linearly spaced range of luminance [0.5,~120]~$cd/m^2$ on (linearly spaced) backgrounds of luminance in the range [1,~160]~$cd/m^2$. These stimuli are easily generated in digital levels (i.e. ready to feed conventional artificial models) with the code provided in this work\footnote{See the script \texttt{WeberCrispening.m}} which makes use of the calibration of the Matlab toolbox Colorlab~\cite{Colorlab02} in a standard computer screen. 

Let's describe the perceived brightness of the stimuli in this test.

\textbf{First}, the series of stimuli in the darkest background clearly show the saturation nonlinearity of the brightness vs luminance curve: note that the jumps in perceived brightness for the low-luminance tests are distinctly bigger 
than the equivalent jumps for the same increments in luminance at the high-luminance end. In the axis of perceived brightness, the above implies that the response (blue curve) has large slope (high sensitivity) at the low-luminance end, and a saturation of such response (lower sensitivity) at the high-luminance end. That makes the \emph{qualitative} saturating blue curve of brightness vs luminance.

\textbf{Second}, when one increases the luminance of the background (e.g. from 1~$cd/m^2$ to 40~$cd/m^2$), the brightness of the (same) samples is lower than in the previous series, so the \emph{qualitative} brightness response to this second series of stimuli is below the previous one (as depicted by the \emph{qualitative} black curve). 

\textbf{Third}, by looking at the stimuli highlighted in gray in the 1~$cd/m^2$ and the 40~$cd/m^2$ backgrounds it is obvious that in the brighter background the stimuli with equivalent brightness are shifted to the right in the scale of luminance, which means that the response in black (for the stimuli in the brighter background) is shifted right-down with regard to the curve in blue (for the stimuli in the darker background). Moreover, this means that the black curve has sigmoidal shape as it should start from zero brightness.
Similar visual reasoning implies that this shift progressively increases as one increases the luminance of the background, as \emph{qualitatively} illustrated by the samples highlighted in gray. 

\textbf{Fourth}, the (same) stimuli in the brightest background in elicit a brightness response with substantially different shape: the sigmoid has substantially shifted to the right (red curve) and, all in all, one can see a smooth transition of the sigmoidal response curves from the blue curve to the red curve. The crispening effect (increased sensitivity around backgrounds of similar luminance) is illustrated by the shift to the right of the points of maximum slope in the response curves.

Finally, \textbf{fifth}, the decreasing brightness of the samples of the same luminance in backgrounds of progressively bigger luminance (as illustrated by the samples highlighted in green) illustrate brightness induction~\cite{Fairchild13}.

Of course the \emph{qualitative} visual observations done here, by no means try to substitute the rich \emph{quantitative} literature in which these responses are determined by accurate psychophysics~\cite{Stiles00,Fairchild13}.
However, (1) the phenomena are compelling enough so that one can see the qualitative trends of the curves by the eye, and, as seen in the numerical experiments below, (2) these trends (visible in ready to use digital images) are enough to spot divergences with human behavior in certain artificial model or discriminate between models in terms of their similarity to human behavior, which is the ultimate goal of the tests presented here. 

\subsubsection{Nonlinear response to saturation and color adaptation}

Responses to constant deviations from white in the red-green and yellow-blue directions of the Jameson \& Hurvich color space~\cite{Jameson57,Vila23} with equiluminant stimuli describe the nonlinear perception of hue and saturation as pointed out in~\cite{Gegen92,Hita93} in similar opponent spaces, i.e. the chromatic version of fact~2 in Table~\ref{tabla1}. 

\begin{figure}[b]
\begin{center}
\vspace{-0.3cm}
\hspace{-2.45cm}\includegraphics[width=1.18\textwidth]{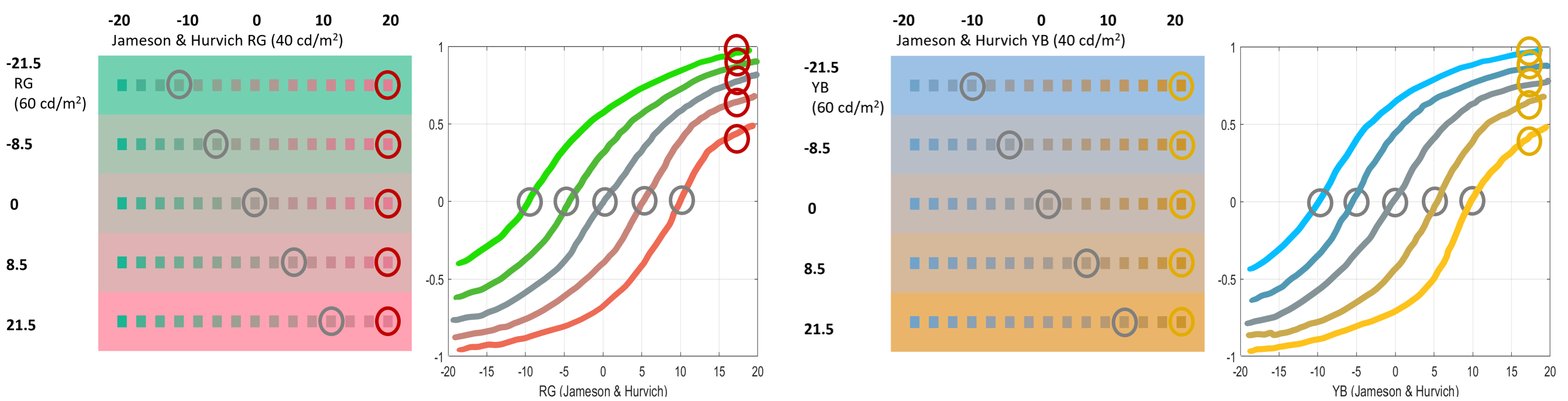}
\vspace{-0.3cm}
\caption{\small{Series of nonlinear perceived saturation (or response of the opponent chromatic channels) versus linearly spaced increments in colorimetrically calibrated color opponent directions in different chromati contexts. This illustrates the nonlinear effects pointed out in~\cite{Gegen92,Hita93}.}}
\label{gegen}
\vspace{-0.9cm}
\end{center}
\end{figure}

Figure~\ref{gegen} shows colorimetrically calibrated stimuli with such deviations (in the range [-20, 20] of the linear RG and YB tristimulus values of the Jameson and Hurvich space) in different backgrounds, which are easy to generate and modify by using the code provided in this work\footnote{See the script \texttt{Gegenfurtner.m} of this work which also uses the Toolbox Colorlab~\cite{Colorlab02} for calibration.}. As in the previous test, let's describe the perceived hue and saturation of the stimuli to infer the qualitative shape of the responses.

\textbf{First}, take the stimuli in gray backgrounds and note that the jumps in perceived hue are bigger around the central (achromatic) stimuli than in the extremes with more saturated stimuli (either red, green, yellow or blue): judge the jumps in saturation  close to the achromatic stimulus and at the extremes of the chromatic axes.
Similarly to the responses for brightness, these differences imply a sigmoidal response to saturation when the stimuli linearly depart from white in constant steps: see the qualitative responses in gray for both the red-green and the yellow-blue directions.
\textbf{Second}, these sigmoidal responses shift to the right or to the left as can be seen from the shift of the stimuli that are perceived as achromatic in the different backgrounds (e.g. see the stimuli highlighted in gray). 
Note that a stimulus is seen as achromatic when the response of the mechanism tuned to red-green or yellow-blue is zero. See the corresponding shifts in the zero crossings of the sigmoids (also highlighted in gray).
Finally, \textbf{third}, the shift of the responses is bigger as the saturation of the background is increased. 

Again, the goal of this test is not substituting the original accurate psychophysics done on humans~\cite{Gegen92,Hita93} to point out these phenomena. On the contrary, they just represent an easy way to get digital images that can be used to test artificial models and check if their responses qualitatively behave like humans.  

\subsubsection{Texture masking 1 (energy): nonlinear adaptive contrast response}

\begin{figure}[b]
\begin{center}
\vspace{-0.3cm}
\includegraphics[width=1\textwidth]{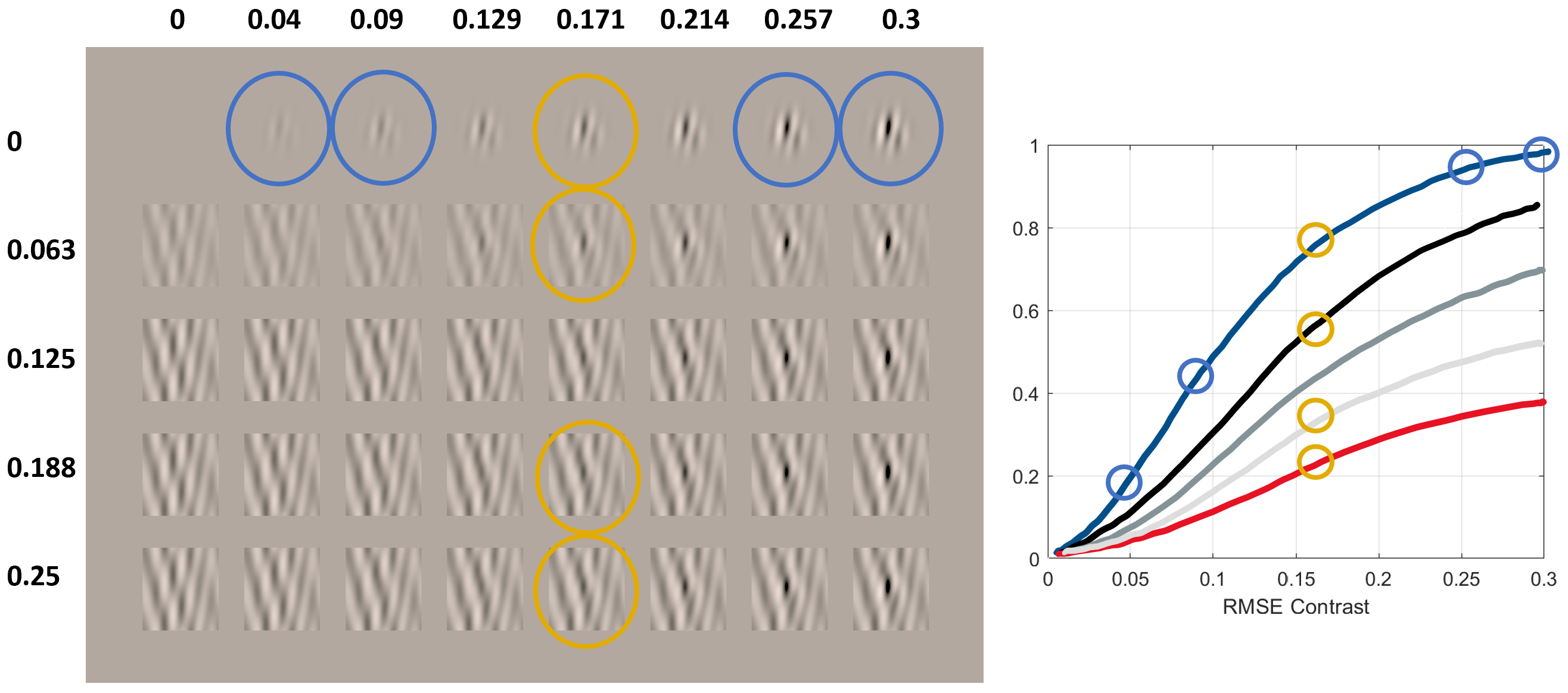}
\vspace{-0.3cm}
\caption{\small{Series of nonlinearly perceived contrast (or response of the mechanisms tuned to certain texture) versus linearly spaced increments in contrast calibrated test of controlled spatial frequency in backgrounds of different (controlled) energy. This illustrates the nonlinear effects pointed out in~\cite{Georgeson75,Legge80,Legge81,Daly90}.}}
\label{masking1}
\vspace{-0.9cm}
\end{center}
\end{figure}

The same kind of qualitative derivation of human-like responses can be applied to the perceived contrast of textured patterns with calibrated frequency content and controlled luminance. 
The test presented here illustrates the fact that perceived contrast nonlinearly depends with linearly increasing Michelson contrast~\cite{Legge80,Legge81} and this response decreases with (is masked by) the energy of a background of similar texture~\cite{Foley94,Watson97}.
This corresponds to fact~8 in Table~\ref{tabla1}.
The stimuli presented in the following example can be reproduced and modified both in frequency orientation, average luminance and contrast with the code provided\footnote{See the script \texttt{MaskingEnergy.m} which makes extensive use of the Toolbox Vistalab~\cite{Vistalab}.}.

Figure~\ref{masking1} shows Gaussian windowed test noise patches of 4 cycles/degree (cpd) in images subtending 1 degree with average luminance of 50 $cd/m^2$ and linearly spaced RMSE contrasts (from left to right) in the range [0, 0.3].
The different rows show the same tests on different backgrounds of noise of 4 cpd with linearly spaced RMSE contrast in the range [0, 0.25].

Similarly as in the previous cases, lets describe the perceived contrast along the two dimensions of the panel. Test: left to right, and background: top to bottom. Again, the qualitative shape of the responses will be determined by the perceived jumps of contrast of the tests (from left to right) and by their variation as one increases the energy of the background (from top to bottom).

\textbf{First}, for the zero contrast background (first, top row) the jumps in perceived contrast in the low-contrast end (left) are bigger than the jumps in perceived contrast in the high-contrast end (right). See the differences in perceived contrast in the tests highlighted in blue. This implies a saturating contrast response curve (as in the previous examples), i.e. the blue curve.

\textbf{Second}, as the contrast of the background is increased (see stimuli highlighted in orange) the perceived contrast of the test is reduced. This implies that subsequent curves (black and lighter shades of gray) are below the initial blue curve.

Finally, \textbf{third}, as in order to perceive the tests with equivalent contrasts in backgrounds of progressively bigger energy the necessary contrast of the test increases, this means that sigmoidal curves shift to the right.

As in the previous examples, the qualitative behavior illustrated by this series of digital images generated by our code should give (in artificial models) corresponding saturating curves with smooth variation from the blue (zero contrast background) condition to the red (high contrast background) condition, and hence the lower response curve.  

\subsubsection{Texture masking 2 (features): interaction between orientations}

\begin{figure}[b]
\begin{center}
\vspace{-0.3cm}
\includegraphics[width=1\textwidth]{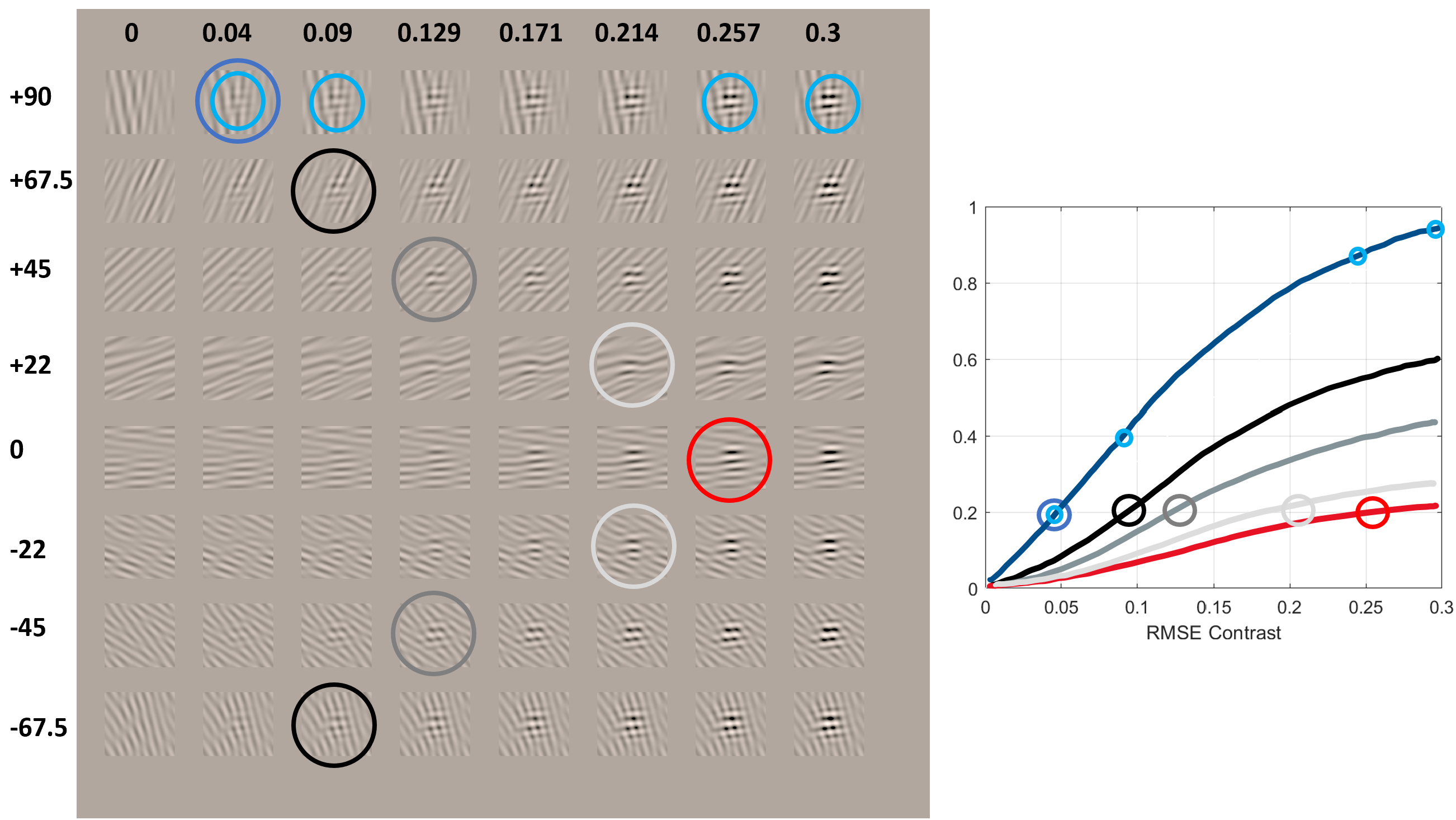}
\vspace{-0.3cm}
\caption{\small{Series of nonlinearly perceived contrast (or response of the mechanisms tuned to certain texture) versus linearly spaced increments in contrast calibrated test of controlled spatial frequency in backgrounds of different orientation. This illustrates the nonlinear effects pointed out in~\cite{Foley94,Watson97}.}}
\label{masking2}
\vspace{-0.9cm}
\end{center}
\end{figure}

Reduction of sensitivity (the so called masking) also happens when certain test is presented on top of a background that shares some feature with the test~\cite{Ross91,Foley94,Watson97}, i.e. facts~9 and~10 in Table~\ref{tabla1}. The next example, Fig.~\ref{masking2}, refers to the specific case of interaction between orientations of test and background. It can be reproduced and modified both in frequency, orientation, contrast and average luminance with the code provided\footnote{See the script \texttt{MaskingOrient.m} which makes extensive use of the Toolbox Vistalab~\cite{Vistalab}.}.

Figure~\ref{masking2} shows 6 cpd horizontal Gabor patches with average luminance of 50 $cd/m^2$ and RMSE contrast increasing linearly from left to right in the range [0,~0.3]. These Gabor patches are shown on top of band-pass noise of contrast 0.2, with the same frequency, but different orientation.
The numbers in the different rows shows the angular difference between tests and backgrounds. 
The figure shows some compelling facts that lead to clear qualitative trends in the response curves. 

\textbf{First}, the test is better seen (has bigger visibility or perceived contrast) when the background is orthogonal to the test (in the first row). In that row the different jumps in visibility in the low-contrast and high-contrast ends (tests highlighted in cyan) indicate a saturating response as in the previous examples (response curve in blue).  

\textbf{Second}, the necessary contrast to detect the test smoothly increases as the difference in orientation between test and background decreases:
see that the tests highlighted in blue, black, shades of gray and red, approximately have the same visibility over the different backgrounds with angular differences in the range [90,0] deg.
The trend is similar for negative angular differences. 
This implies a smooth variation (decrease) of the response curves in terms of the difference between test and background.

\textbf{Third}, the biggest masking is obtained when test and background are aligned (the red curve is clearly the lowest response curve). This and the previous fact imply that the general trend is this smooth transition of the nonlinear curves from the blue to the red. 

\subsection{Proposed methodology}

\begin{figure}[b]
\begin{center}
\vspace{-0.3cm}
\includegraphics[width=0.8\textwidth]{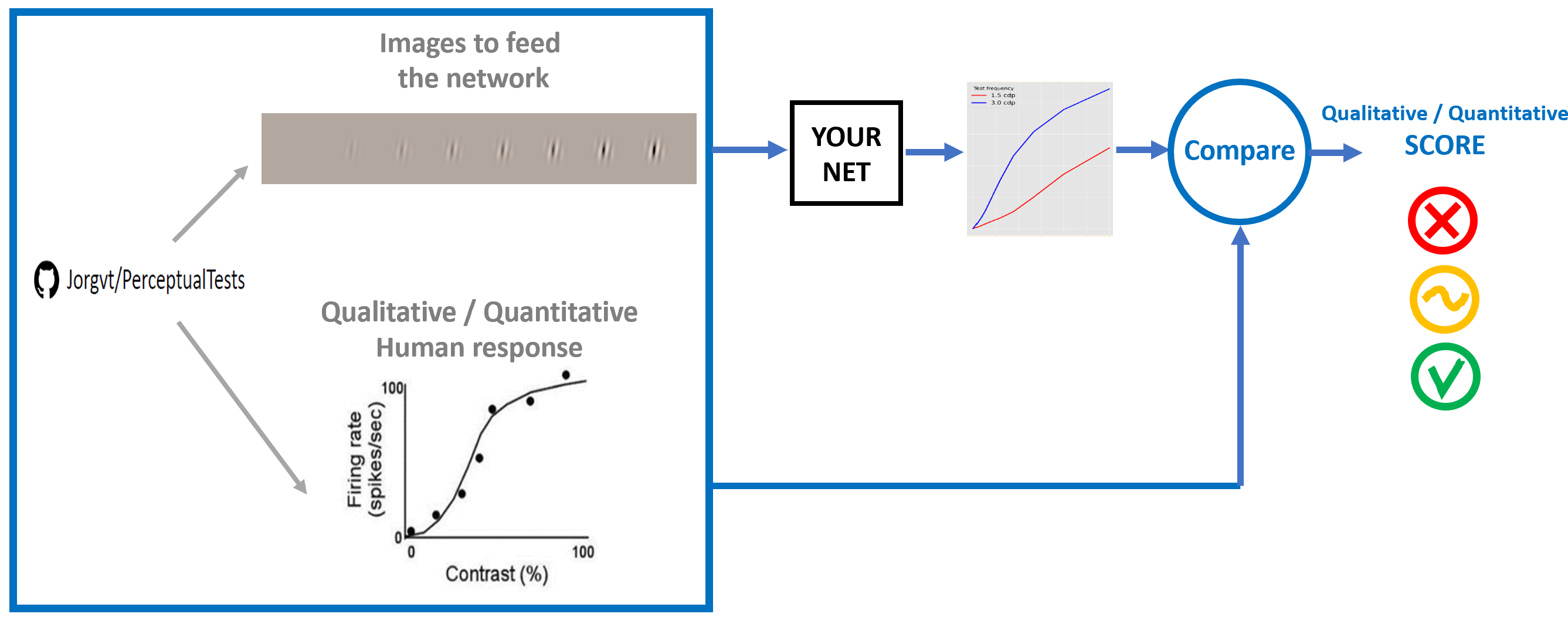}
\vspace{-0.3cm}
\caption{\small{The proposed method: feed the model with series of images, compute responses (using the simplest possible read-out mechanism) and make quantitative comparisons with standard sensitivity surfaces or qualitative comparisons checking the nonlinearitiy using different adaptation conditions.}}
\label{method}
\vspace{-0.9cm}
\end{center}
\end{figure}

Our proposal is simple: just use the code provided here\footnote{The Decalogue Toolbox is available here: \texttt{http://isp.uv.es/docs/TuringTestVision.zip}} to generate the stimuli (digital images well calibrated in luminance, color and spatio-temporal frequency) that illustrate the 10~compelling facts listed in Table~\ref{tabla1} that describe the adaptive information bottleneck of low-level human vision. The resulting digital images are organized in series that correspond to progressive stimulation of a vision system in particular ways.
The interesting point is that this set of controlled stimulation conditions lead to intuitive responses (as shown above), or even to standardized sensitivity curves or surfaces that are also provided with the code.

Once stimuli are generated they are used to feed any artificial image-computable model. Then, depending on the model, the user decides where to read from the network under consideration and the read-out mechanism to get \emph{visibility} values to generate artificial series of response curves.

In the case of facts~2, 8, 9 and~10 these curves have to be compared with the kind of qualitative curves described above, which given the clarity of the selected stimuli can be drawn by simple visual observation of the stimuli as described above. 

In the case of fact~5 (existence of center-surround and Gabor-like receptive fields tuned to achromatic, red-green and yellow-blue patterns~\cite{Shapley11}), the more straightforward method is checking their presence by reading the response to deltas from single neurons or from the Jacobian of the network at that layer~\cite{Martinez18,GomezVilla20,Li22}. Other indirect methods could be (1)~using reverse correlation feeding the network with controlled noise (also generable using Vistalab~\cite{Vistalab} following the appropriate literature~\cite{Eckstein02}), or (2)~using artificial psychophysics based on adaptation (e.g. the Blakemore and Campbell experiment~\cite{Blakemore69}). However, this very last method to measure fact~5 relies on fulfillment of adaptation facts 6-10, which may not hold in non-human networks.

In the above (non-standardized) cases the general trends of the curves can be qualitatively assessed in detail: general shape of the curves, the blue response and the red curve being the biggest and the lowest respectively, and the transition from one to the other. Note that user of the provided code can change the parameters of the stimuli and infer new curves by applying a similar visual analysis. For the receptive fields they can be analyzed using shape parameters in the spatial or the Fourier domain as classically done in visual neuroscience~\cite{Ringach02,Ringach04,Martinez17,Loxley17}
and the same for the chromatic tuning in standard color spaces~\cite{Lennie08,Gutmann14,GomezVilla20}.

Finally in the case of sensitivity curves or surfaces which are standardized or available in the code (facts 1, 3, 4, 6 and 7) the visibility values obtained from the models can be numerically compared with the provided ground truth.

This qualitative/quantitative methodology is summarized in Fig.~\ref{method}, and applied in the next experimental section for three illustrative networks.


\section{Experiments: analysis of three illustrative deep-models}

\subsection{Networks and experimental setting}

In our experiments we check the behavior on the proposed \emph{Decalogue} of three recent networks of similar architecture:
\begin{enumerate}
\item A parametric vision model, the \textbf{\emph{BioMultiLayer}} network~\cite{Martinez18}, which consists on a cascade of four linear+nonlinear stages that account for 
(1)~color opponency and adaptation, 
(2)~contrast computation, 
(3)~contrast sensitivities and energy masking, and 
(4)~wavelet analysis and cross-masking between textures.   
The linear parts of all the stages were not optimized but they were directly inspired by classical psychophysical or physiological literature. 
The nonlinear parts were implemented via Divisive Normalization~\cite{Carandini94,Malo06a,Laparra10,Malo24,Carandini12}. The nonlinearities of the 2nd and 3rd stages of the model were tuned via the psychophysical method of Maximum Differentiation in~\cite{Malo15}.
And the nonlinear parts of the 1st and 4th stages were tuned to reproduce subjective opinions on distortion and contrast masking facts~\cite{Martinez18,Martinez19}. The statistical properties of the model and its relations with recurrent models were studied in~\cite{Gomez19} and~\cite{Malo24} respectively.
\vspace{0.1cm}

\item A non-parametric model to predict subjective image quality, the \textbf{\emph{PerceptNet}}~\cite{Hepburn20}, which
starts with a nonlinear front-end at the retina followed by a cascade of three linear+nonlinear stages. The architecture was intended to accommodate similar vision facts that motivated the \emph{BioMultiLayer}. The \emph{PerceptNet} architecture is similar to AlexNet~\cite{Alexnet} but its nonlinearities were formulated using an end-to-end optimizable Divisive Normalization~\cite{Laparra17,Balle17}.    
Both the linear and the nonlinear parts of \emph{PerceptNet} were end-to-end tuned to maximize the correlation with humans on subjective image distortions~\cite{Hepburn20}. Non-parametric layers of \emph{PerceptNet} are not easy to interpret as pointed out recently~\cite{Vila25b}. 
\vspace{0.1cm}

\item An image segmentation model, the \textbf{\emph{Bio~U-Net}}~\cite{Hernandez23}, with the same style encoder as the nonparametric \emph{PerceptNet} (a cascade of linear + divisive normalization stages), but augmented with a decoder that recovers the original dimension of the input signal and predicts a class per pixel for semantic segmentation. The encoder and the decoder were tuned to optimize segmentation in different databases. The benefits of the biologically-inspired nonlinearities of this model for segmentation have been further studied thereafter~\cite{Hernandez24}.

\end{enumerate}

We assumed a visual field of 2 degrees with a sampling frequency of 64 cycles/deg, i.e. we fed the models with $128\times128$ images. We measured the responses of the models to specific tests through the Euclidean departure between the response to test+background with regard to the response to the isolated background. 

\subsection{Results}

\subsubsection{Spectral sensitivities and color responses (properties 1 and 2)}

\begin{figure}[t]
\begin{center}
\vspace{-0.3cm}
\hspace{-1.5cm}\includegraphics[width=1.1\textwidth,height=9.5cm]{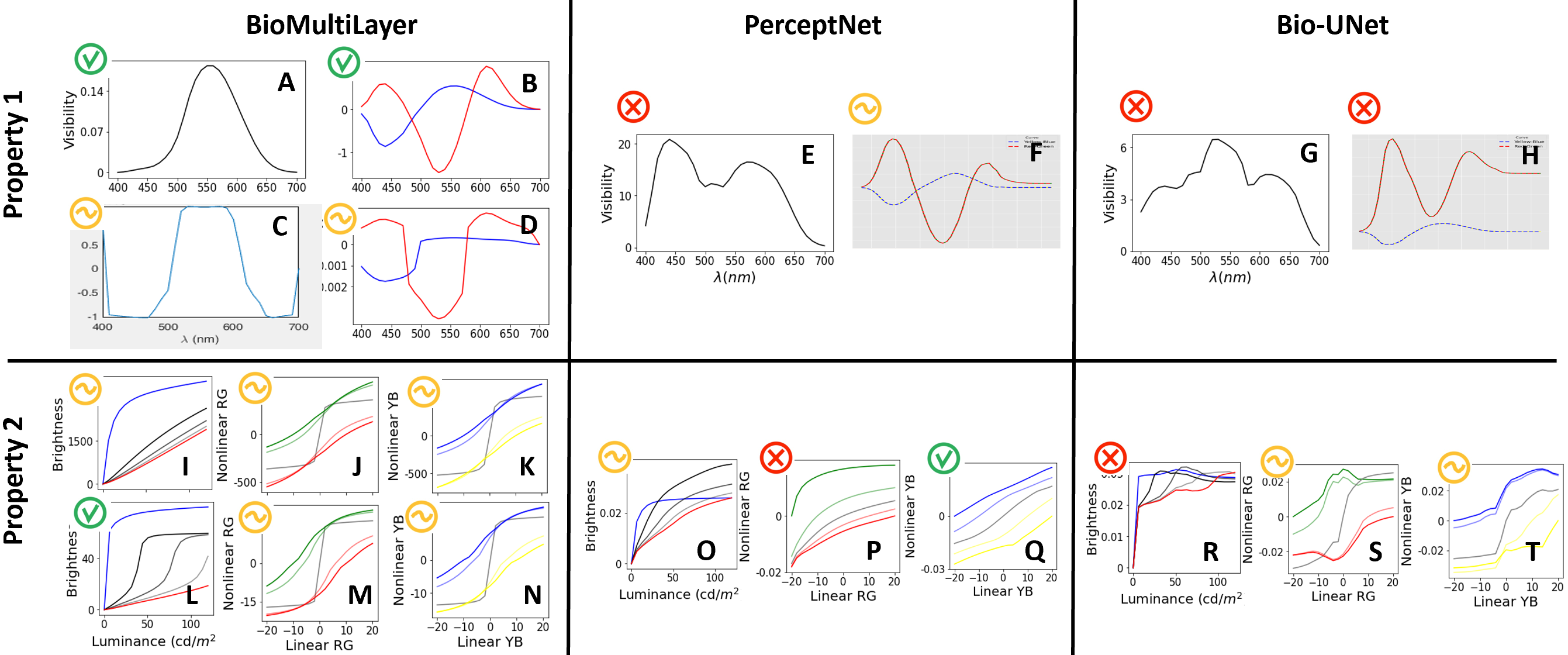}
\vspace{-0.2cm}
\caption{\small{Spectral sensitivities of the considered models (top) and corresponding responses to 
luminance and linear deviations from white 
 in the cardinal red-green and yellow-blue directions (bottom). On the one hand, knowledge of the standard spectral sensitivity, the CIE $V_\lambda$ curve~\cite{Stiles00}, or the standard spectral sensitivty of the opponent channels~\cite{Jameson57} indicates which model is trivially correct in Prop.~1. On the other hand, the stimuli proposed here (Figs.~\ref{weber} and~\ref{gegen}) and the associated human behavior described above indicates the correct trends in the responses of Prop.~2.}}
\label{color_results}
\vspace{-0.5cm}
\end{center}
\end{figure}

Figure~\ref{color_results}-top shows the response of the models to quasi-monochromatic stimuli\footnote{Spectrally narrow Gaussians (5 nm width) of constant energy centered on different wavelengths along the visible spectrum on top of a low energy flat spectrum, as in Fig.~\ref{spectral}. In this way all the stimuli can be faithfully represented in digital values.} to get the spectral sensitivity of the neurons (property~1). 
In order to point out the relevance of the layer where responses are measured from, in the case of the \emph{BioMultiLayer} network, we consider direct read out of the response (with sign) in the first linear layer (subplots A and B) and in the last nonlinear layer (subplots C and D).
In this network the first linear layer has achromatic and opponent channels defined by construction so
the $V_\lambda$~\cite{Stiles00} (subplot A) and the opponent curves of Jameson $\&$ Hurvich~\cite{Jameson57} (subplot B) are trivially obtained. Interestingly, the spectral sensitivities at the last nonlinear layer are wide-band positive in the first channel of the network and opponent in the other two channels, but their shapes are substantially modified with regard to the human-like behavior at the first layer. These differences justify the qualitative scores given in each case. We can conclude that spectral sensitivity in this model is human-like at the front-end but degrades throughout the network. In other words, as suggested in section 2-b, read-out location matters and certain kind of information should be extracted from a specific place of the model.

The \emph{PerceptNet} has a color space change after the retinal nonlinearity. There is where we measure the sensitivities as the design idea was that achromatic and chromatic channels emerged at that stage. Results show that the first channel displays an all-positive but bimodal response (subplot E) and the other two channels display opponent-like responses (subplot F). 

The very same location of the encoder of the segmentation \emph{Bio-U-Net} has very different sensitivities despite it has the same architecture as the \emph{PerceptNet} up to that layer. The sensitivity of the (supposedly) achromatic channel is very noisy and the other two channels display qualitatively opponent oscillations but they are shifted in absolute value (subplots G and H).  

On the other hand, Fig.~\ref{color_results}-bottom checks property~2 by showing the responses to (i)~luminance and to deviations from white in the (ii)~red-green and (iii)~yellow-blue directions (left, center and right respectively).
In the achromatic case, tests in the range [0.5, 120]~$cd/m^2$ are shown on top of backgrounds of different luminance in the range [1, 160]~$cd/m^2$. The response curves in different backgrounds are depicted in blue, black, and progressively lighter shades of gray until red, as in Fig.~\ref{weber}. 
In the chromatic cases, responses are computed with tests on an achromatic background (black curve) and on backgrounds of progressively saturated color (reddish and greenish curves and blueish and yellowish curves as in Fig.~\ref{gegen}).

For the \emph{BioMultiLayer} model we have such responses for two different layers: first (I, J, K) and fourth (L, M and N).
The achromatic response of the first layer is certainly nonlinear for the darkest background, and the response gets attenuated when the luminance of the background is increased (see the transition from curves in blue to red in subplot I). However, these responses do not reproduce the crispening (sigmoids shifting to high luminance), and responses for high luminance backgrounds are too linear. As a result the achromatic behavior of this layer has been qualified as non-human. 
The chromatic responses display sigmoidal shape and they shift in the right directions under different backgrounds (subplots J and K). However, the nonlinearities for the chromatic backgrounds are very smooth compared to the sharpness of the nonlinearity for the achromatic background. As a result, the human similarity of chromatic behaviors have been qualified as intermediate.
In contrast, the achromatic response of the 4th layer (subplot L) does reproduce the nonlinear behavior and crispening, so it has been qualified as more human-like than the achromatic response of the 1st layer. 
Shifts of the chromatic nonlinearities are stronger depending on the background, but the nonlinearities in achromatic backgrounds (black curves in subplots M and N) are still too sharp. Therefore the score remains the same.

The \emph{PerceptNet} model displays nonlinear behavior and crispening in the responses to the achromatic series (subplot O). However, Note how the curves corresponding to light backgrounds exceed the response on dark backgrounds, so human similarity has been qualified as intermediate.
The responses to red-green series in \emph{PerceptNet} shift in the right directions on different backgrounds, but they are too linear (and hence wrong) in subplot P. In contrast, the blue-yellow responses (subplot Q) display a rather human behavior.

Finally, the \emph{Bio-U-Net} shows a clearly non-human achromatic response: note the noise and wrong order in the curves with no trace of crispening (subplot R).
In contrast, the responses to the chromatic series display the expected sigmoidal shape with the shift in the proper directions for the different chromatic backgrounds (subplots S and T). Noisy and unstable responses is what determined the intermediate score.

\subsubsection{Achromatic and chromatic contrast sensitivities and receptive fields \\  (properties 3, 4 and 5)}

\begin{figure}[t]
\begin{center}
\vspace{-0.3cm}
\hspace{-2cm}\includegraphics[width=1.17\textwidth]{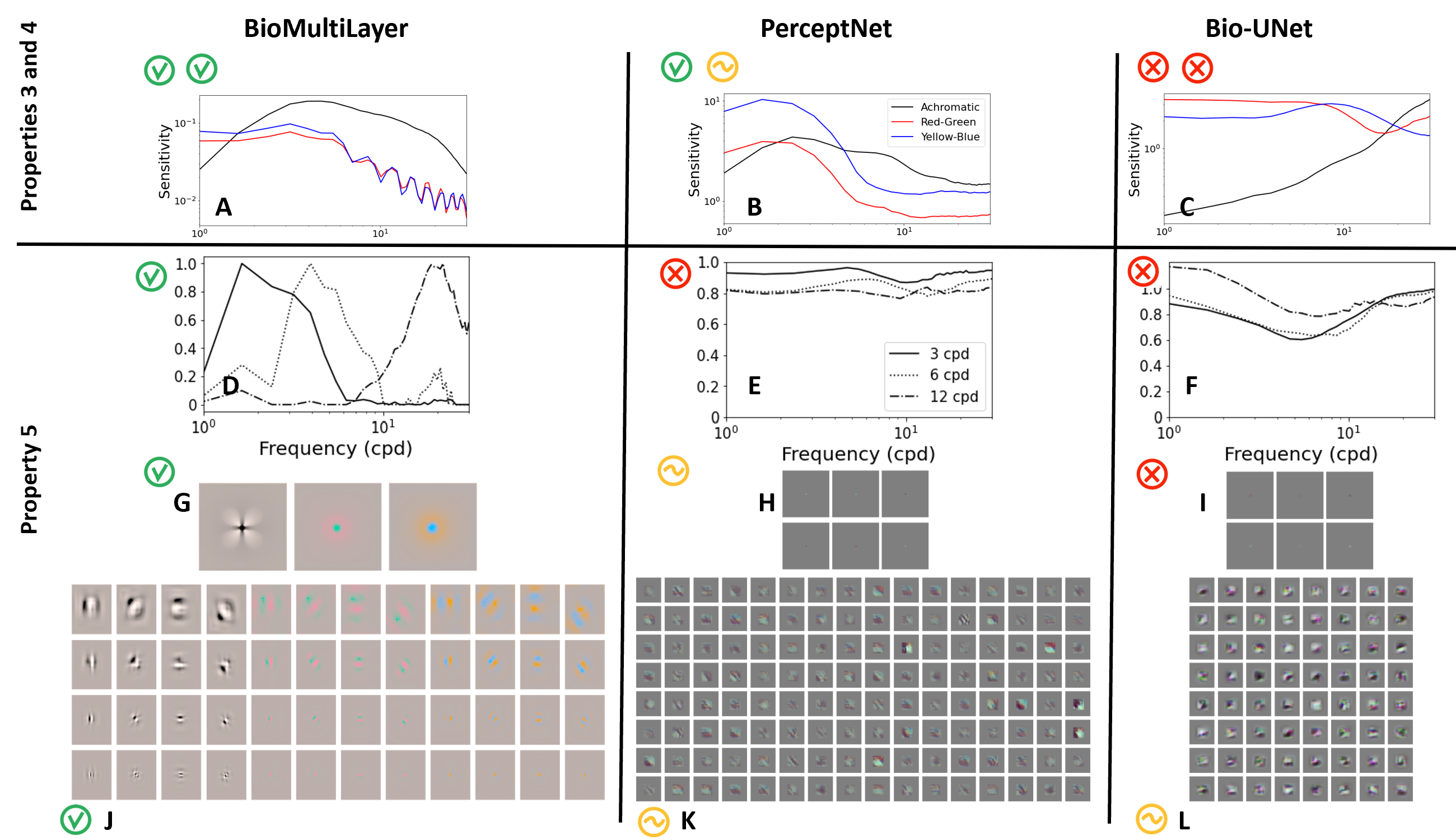}
\vspace{-0.3cm}
\caption{\small{Achromatic and chromatic contrast sensitivities of the considered models (top subplots A, B, C), and different sets of receptive fields computed in different ways at different depths of the models (bottom). See text for details on the psychophysical and the physiological ways to estimate the receptive fields.}}
\label{RF_CSF_results}
\vspace{-0.4cm}
\end{center}
\end{figure}

The top row of Fig.~\ref{RF_CSF_results} shows the achromatic Contrast Sensitivity Function (property~3, black curve) and the red-green and yellow-blue Contrast Sensitivity Functions (property~4, red and blue curves respectively). 
These CSFs have been computed from the responses 
to noise patterns of controlled spatial frequency and the same low contrast ($C_{\textrm{RMSE}}=0.05$) for every frequency. 
Patterns were generated in the corresponding color channel of the Jameson \& Hurvich color space~\cite{Jameson57} that isolate luminance, red-green, yellow-blue components. 
We consider the responses at the last layer of the networks and we plot the Euclidean distance between the responses for each pattern and for a flat image of the same average color.

The CSFs of the \emph{BioMultiLayer} model (subplot A) strongly resemble the human CSFs~\cite{Campbell68,Mullen85}: the achromatic response is band-pass with peak sensitivity around 4~cpd and high cut-off frequency (above 32 cpd), and the chromatic responses are lower and basically low-pass with cut-off frequencies about 15 cpd.

The achromatic CSF of the \emph{PerceptNet} is also band-pass (black curve in subplot B), but the chromatic CSFs are far from human because their shape is also bandapass and the responses to modulations in the YB direction are way bigger than the responses to equivalent achromatic modulations.

Finally, the \emph{Bio-U-Net} (subplot C) displays a strongly non-human behavior: see the non-plausible high-pass behavior of the responses to achromatic gratings, and the bigger responses to chromatic gratings in the mid-frequency range.  


The receptive fields of the models (property~5) have been proved in two ways: (1)~a \emph{psychophysical} method based on the Blakemore \& Campbell experiment~\cite{Blakemore69}, which relies on the attenuation of the CSF under adaptation for different frequencies, and (2)~a \emph{physiological} method based on recording the response to deltas in the luminance, red-green and yellow-blue channels~\cite{Martinez18}.  
In the \emph{BioMultiLayer} model the attenuation of the achromatic CSF when the gratings are shown on top of backgrounds of specific frequencies (subplot D) reveals the existence of narrow-band sensors with bandwidth that increases with frequency, which is consistent with human behavior~\cite{Blakemore69,Simoncelli90}.
This comes from the fact that the linear part of the 4th layer of this model is made of wavelet kernels and their response is nonlinearly attenuated by the activity of neighbor sensors tuned to the same feature through Divisive Normalization. 

On the other hand, when checking the shape of the receptive fields using delta functions one gets two biologically plausible results:
(a)~in the 3rd layer of the \emph{BioMultiLayer} network receptive fields are center-surround patterns in the achromatic, red-green and yellow-blue directions (subplot G), and (b)~in the 4th layer one gets local frequency filters with different orientations and scales (subplot J) as happens in biological vision at LGN~\cite{DeAngelis97,Shapley11} and V1~\cite{Hubel59,Watson83}. 


\begin{figure}[t]
\begin{center}
\vspace{-0.3cm}
\hspace{-1.5cm}\includegraphics[width=1.15\textwidth,height=10.5cm]{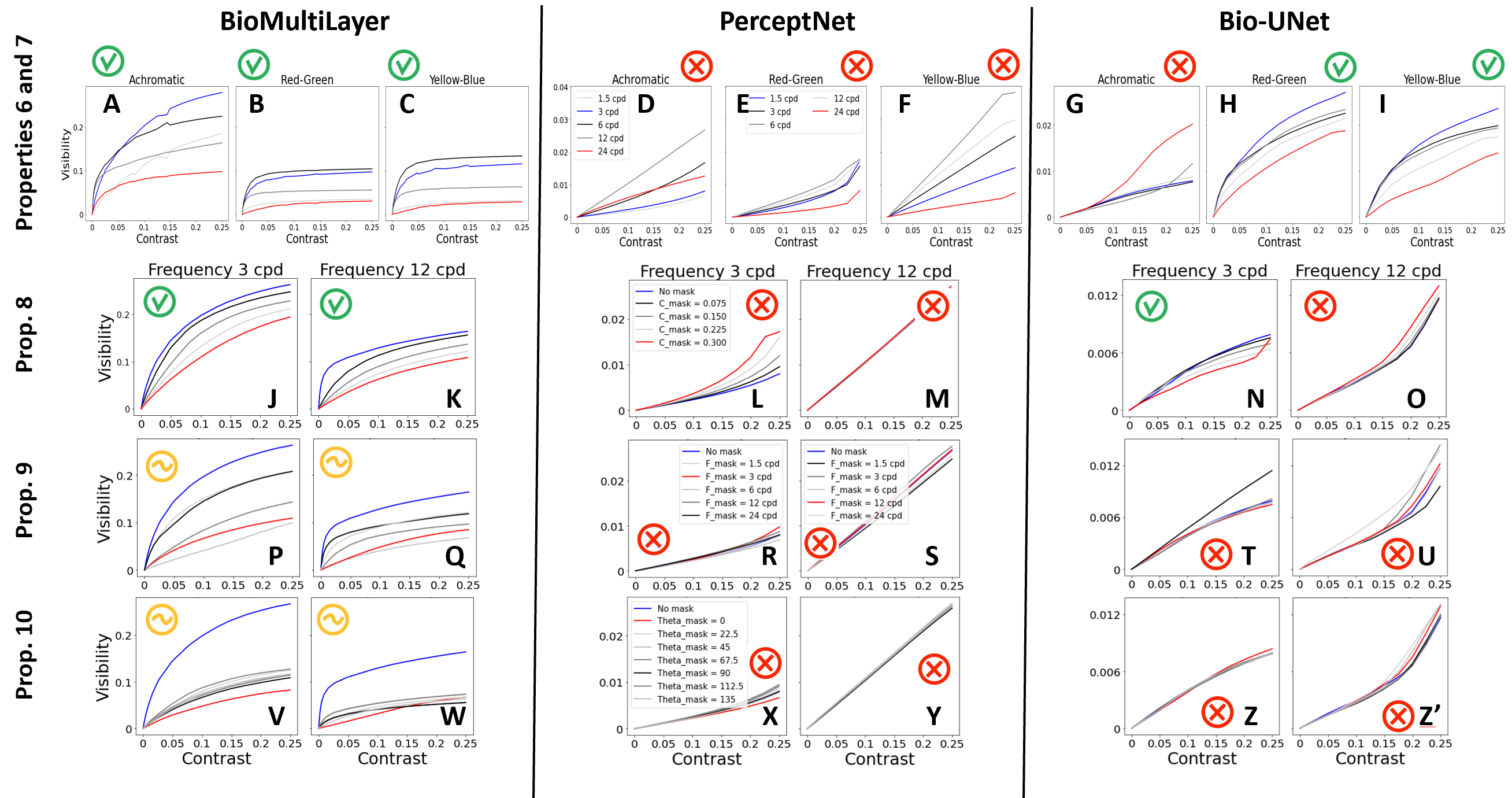}
\vspace{-0.3cm}
\caption{\small{Contrast responses of the considered models in different masking conditions and for achromatic and chromatic textures of different frequencies. The color code (indicated in the subplots corresponding to Percepnet, but applicable to the equivalent curves of the other models) has been designed so that human response curves would be in the blue-red order, as in Figs.~\ref{masking1} and~\ref{masking2}. In this way it is obvious which model reproduces better the human behavior.}}
\label{masking_results}
\vspace{-0.5cm}
\end{center}
\end{figure}

For \emph{PerceptNet} results are quite different: first, the Blakemore and Campbell experiment only shows filters tuned for low frequencies (subplot E) which does not correspond to human behavior. This is not only due to the non-human nature of the CSFs, it also means that any frequency leads to a strong attenuation of the responses to high frequency patterns. 
This departure from human behavior is also visible when getting the receptive fields from the last layer of the network using deltas: one gets oriented filters but all with the same size and with low-frequency blobs. More over, chromatic information is spread along all the filters (subplot K), as opposed to what happens in the early layers where one does get achromatic, red-green and yellow-blue responses (subplot H). 

Finally, in the \emph{Bio-U-Net} model, the Blakemore and Campbell experiment also leads to non-human ratios of the CSFs (subplot F). The receptive fields obtained from deltas in the first layers lead to center-surround blobs despite not in definite chromatic directions (subplot I). In the central layers of the encoder one gets larger receptive fields which have no clear spatial oscillations nor preferred chromatic directions (subplot L).

\subsubsection{Contrast saturation, dependence with frequency (properties 6 and 7)}

The top row of Fig.~\ref{masking_results} shows the visibility response (Euclidean difference of response wrt the response of a uniform gray image) for achromatic patterns of different frequencies and for red-green and yellow-blue patterns of different frequencies all seen in isolation (properties 6 and 7). Line styles for the different frequencies in the achromatic and the chromatic cases is different according to the different order expected for band-pass and low-pass systems. In every case, match with human behavior would be illustrated by having the blue curve at the top and the red curve at the bottom, with a smooth transition from black to light gray, in between.

The \emph{BioMultiLayer} model leads to saturating responses  with larger intensities in the achromatic case (left) than in the chromatic cases (subplots A and B-C), as in humans~\cite{Watson97,Uriegas97}.
The achromatic response to mid-frequency (3 cpd, in blue) is clearly bigger than the response to the other frequencies, which is smoothly reduced for higher frequencies (from 6 to 24 cpd) and also attenuated for 1.5cpd. 
On the other hand, the chromatic responses are basically ordered according to frequency in a low-pass fashion. All these trends are in good agreement with the human behavior.

The achromatic responses of the \emph{PerceptNet}, though band-pass, exhibit a quite linear, non-saturating or even expanding, behavior (subplot D). Moreover, these achromatic responses are not bigger than the response to chromatic patterns, particularly the yellow-blue (subplot F), which is contrary to human perception.  

The responses for the chromatic patterns in the \emph{Bio-U-Net} model exhibit human-like saturation and they are in the right frequency order (subplots H and I), but they are larger than the responses for achromatic patterns (subplot G), which is contrary to human perception.

\subsubsection{Energy masking and feature masking (properties 8-10)}

Each panel of the second row in Fig.~\ref{masking_results} shows the responses to a 3 cpd achromatic pattern (left) and a 12 cpd achromatic pattern (right) seen on top of a masking pattern (noise of the same frequency and orientation) with progressively larger RMSE contrast (in the range [0,0.3]) leading to different response curves in different color (from blue to red), thus checking the effect of the energy of the background (property~8). The color code has been selected so that the no-mask case is depicted in blue (less attenuated in humans) and colors from black to light-gray and red are taken for progressively bigger contrasts of the mask.

The responses of the \emph{BioMultiLayer} in Fig.~\ref{masking_results} (subplots J,K) progressively attenuate as the energy of the background is increased in line with the reduction in visibility of the test shown in each column of Fig.~\ref{masking1}. And this happens both for low and high frequency, with bigger responses for the mid- frequency. Therefore the behavior is qualitatively human.
The \emph{PerceptNet} displays a completely non-human behavior: for the 3 cpd tests, progressively larger masks induce enhancement of the expansive (non-saturating) response, and the responses for the high frequency patterns are larger, linear and do not show significant variation with the mask.
Finally, the \emph{Bio-U-Net} model does display human-like attenuation of the response to 3 cpd patterns (subplot N). However, the responses to 12 cpd patterns (subplot O) are not human-like because their (large) size and their expansive shape and increase with the energy of the mask. 

The panels of the third row of Fig.~\ref{masking_results} show the responses for an achromatic test of 3 cpd (left) and 12 cpd (right) seen on top of backgrounds of different frequencies (and 0.2 contrast) compared to the no-mask condition, i.e. it checks the frequency cross-masking (property~9). The color code has been selected so that the no-mask case is depicted in blue (less attenuated in humans) and colors from black to light-gray and red are taken for progressively closer frequencies in mask and test, which lead to increased attenuation of response in humans.

The response of the \emph{BioMultiLayer} model is bigger in the no-mask condition, displays substantial attenuation when the background shares the same frequency of the test (red curves in subplots P and Q) and responses are bigger for 3 cpd than for 12 cpd. In each case the optimal frequency is not the one that leads to the bigger attenuation, but it is close to it. 
The \emph{PerceptNet} responses are not human because for the low frequency, subplot R, responses are not-saturating regardless of the mask, and the responses for high frequency are larger, linear and the presence of backgrounds leads to larger responses (subplot S). The \emph{Bio-U-Net} does not show human-like trends because in the case that displays a saturating response the presence of a background leads to responses bigger than in the no-mask case (black curve in subplot T). The behavior in subplot U are non-human for the same reasons stated in the subplots N and O. 

Finally, the last row of Fig.~\ref{masking_results} shows the responses for low- and high-frequency achromatic patterns (left and right, respectively) seen on top of backgrounds of the same frequency but different orientations, i.e. it checks the orientation cross-masking (property~10). Again, the color code has been chosen so that in a human the blue curve would be at top and the red would be at the bottom as in Fig.~\ref{masking2}.

For this last example, the \emph{BioMultiLayer} model gets bigger attenuation for the background of the same orientation, particularly for high frequency (see the red curves), and the other orientations lead to responses that are between the no-mask condition (in blue) and the same-orientation background (in red) in subplots V and W.
The other models give clearly non-human results because (on top of the arguments used in previous cases) stimulation on backgrounds of the same orientation (red curves) do not lead to the expected attenuation, and bigger attenuation is obtained for backgrounds that are almost orthogonal to the test, which is not what humans experience in Fig.~\ref{masking2}.

\subsubsection{Summary of results}

The qualitative evaluation of the considered models over the proposed tests
is summarized in Table~\ref{qualitative_results}. From this table there is a clear ranking of the alignment between the models and humans. It is not surprising that the parametric model (the \emph{BioMultiLayer}) has bigger alignment in the linear parts (properties 1 and 5) since sensitivities and center-surround and Gabor receptive fields were parametrically built in that model model. 

More interestingly, the band-pass behavior of the sensors emerged from modifications in the CSFs in our simulation of the Blakemore and Campbell experiment. It is also interesting the close reproduction of the band-pass and low-pass behavior and the relative scaling of the CSFs obtained from responses to sinusoids (an original check done here) since they were not built in. This indicates that the (non-trivial) gain of the center-surround cells and the Gabor cells was properly adjusted through the indirect psychophysical experiments done to set its parameters. As a result, the relative order of the (saturated) frequency responses (property 7) is also ok, both for achromatic and chromatic textures. The saturation of the responses to Gabor stimuli in isolation (property 6) is better reproduced in the parametric model than in the Bio-UNet. The difference between them is more evident when one digs in using properties 8-10 because they need proper interaction between texture sensors and this was only easy to do in a parametric model such as the \emph{BioMultiLayer}.

However, note that the reproduction of the interaction between features (both in color, property~2, and in texture, properties~9 and~10) is not properly reproduced not even in the \emph{BioMultiLayer} pointing out that more work is needed in adjusting its parameters as discussed below.

According to the proposed test, the other two models (the -non parametric- \emph{PerceptNet} and the \emph{Bio-U-Net}) are \emph{less human}, in that order of alignment. 
This also makes sense because the \emph{PerceptNet} was tuned to reproduce low-level human opinion on distortion, while the \emph{Bio-U-Net} was just tuned to reproduce a specific mid-level vision goal such as image segmentation. In the discussion we elaborate more on the combination of goals that may explain the organization of the visual system. 

In any case, we see that even with this  qualitative application of the proposed test (again, quantitative comparisons could be done with properties 1, 3, 4 and 6, even for moving patterns) a significant ranking  is possible, and, as discussed below, the qualitative behaviors, when they are properly understood suggest significant changes in the architectures and training of the models. Quantitative automation of the optimization should be iteratively done by alternating goals of different nature as suggested in~\cite{Martinez19}: optimize for conventional goals and then fine-tune to reproduce the effects pointed out by the test proposed here (or the other way around). 

\begin{table}[h!]
\begin{center}
\hspace{-0.5cm}\begin{tabular}{ccccc}
  & Facts      & BioMultiLayer & PerceptNet & Bio-UNet \\\hline
1 & Spectral Sensitivities (achromatic and opponent) & {\color{Green}\CheckmarkBold}{\color{Green}\CheckmarkBold} & {\color{red}\tikzxmark}\begin{large}{\color{Orange}$\mathbf{\sim}$}\end{large} & {\color{red}\tikzxmark}{\color{red}\tikzxmark} \\ 
2 & Brightness \& Color Response Saturation & {\color{Green}\CheckmarkBold} \begin{large}{\color{Orange}$\mathbf{\sim}$}\end{large} & \begin{large}{\color{Orange}$\mathbf{\sim}$}\end{large}\begin{large}{\color{Orange}$\mathbf{\sim}$}\end{large} & {\color{red}\tikzxmark}\begin{large}{\color{Orange}$\mathbf{\sim}$}\end{large} \\
3 & Achromatic Contrast Sensitivity (Bandwidth) & {\color{Green}\CheckmarkBold} & {\color{Green}\CheckmarkBold} & {\color{red}\tikzxmark} \\
4 & Chromatic Contrast Sensitivity (Bandwidth) & {\color{Green}\CheckmarkBold} & \begin{large}{\color{Orange}$\mathbf{\sim}$}\end{large} & {\color{red}\tikzxmark} \\
5 & Spatio-Chromatic Receptive Fields & {\color{Green}\CheckmarkBold}{\color{Green}\CheckmarkBold}{\color{Green}\CheckmarkBold} & {\color{red}\tikzxmark}\begin{large}{\color{Orange}$\mathbf{\sim}$}\end{large}\begin{large}{\color{Orange}$\mathbf{\sim}$}\end{large} & {\color{red}\tikzxmark}{\color{red}\tikzxmark}\begin{large}{\color{Orange}$\mathbf{\sim}$}\end{large}\\
6 & Nonlinear Contrast Response: Saturation & {\color{Green}\CheckmarkBold}{\color{Green}\CheckmarkBold} &  {\color{red}\tikzxmark}{\color{red}\tikzxmark} & {\color{red}\tikzxmark}{\color{Green}\CheckmarkBold} \\
7 & Nonlinear Contrast Response: Frequency order & {\color{Green}\CheckmarkBold} & {\color{red}\tikzxmark} & \begin{large}{\color{Orange}$\mathbf{\sim}$}\end{large} \\
8 & Context effects: Energy & {\color{Green}\CheckmarkBold} & {\color{red}\tikzxmark} & \begin{large}{\color{Orange}$\mathbf{\sim}$}\end{large} \\
9 & Context effects: Frequency & \begin{large}{\color{Orange}$\mathbf{\sim}$}\end{large} &  {\color{red}\tikzxmark} & {\color{red}\tikzxmark} \\
10 & Context effects: Orientation &\begin{large}{\color{Orange}$\mathbf{\sim}$}\end{large} &  {\color{red}\tikzxmark} & {\color{red}\tikzxmark} \\
\end{tabular}
\vspace{0.5cm}
\caption{\small{Summary of qualitative results is enough to discriminate between the three models.}}
\label{qualitative_results}
\end{center}
\end{table}
\vspace{-0.5cm}

\section{Discussion: \\ 
What can be learnt from the proposed methodology?}

In this section, we discuss the benefits of the proposed \emph{Decalogue} for generic artificial models.
Benefits go beyond the evaluation of the human nature of models: even if we don't need that certain model is similar to humans, the behaviors described by the human-like curves elicited by the stimuli in the \emph{Decalogue} imply human-like bottlenecks and adaptation properties that one would like in efficient and robust artificial vision systems. Similarly, we also discuss the benefits of the architectures from classical vision science models that reproduce such behaviors.

\subsection{(Non-human) curves suggest changes in the architectures}

\begin{figure}[t]
\begin{center}
\vspace{-0.3cm}
\includegraphics[width=0.9\textwidth]{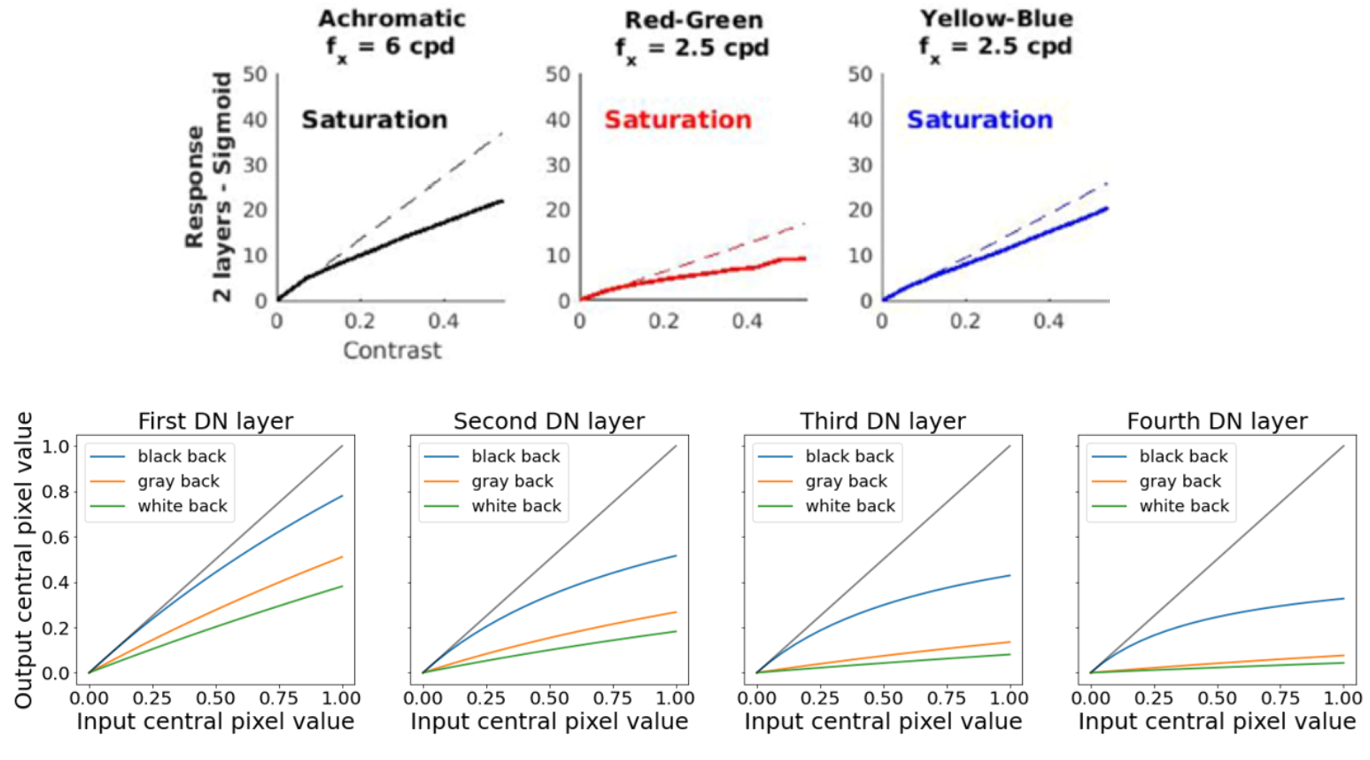}
\vspace{-0.3cm}
\caption{\small{(Top)~Human-like saturation behavior in contrast response happening in generic shallow autoencoders~\cite{Li22}. (Bottom)~Human-like behavior obtained in image segmentation U-Nets when they are equipped with bio-inspired Divisive Normalization to improve their performance~\cite{Hernandez24}.}}
\label{saturation_encoders}
\vspace{-0.3cm}
\end{center}
\end{figure}

When measuring the response of conventional networks using the spatially and chromatically calibrated stimuli 
proposed here one can get human-like behaviors such as the ones shown in Section~3. For instance, shallow autoencoders optimized for image deblurring and denoising display human-like saturation when responding to achromatic and chromatic gratings of controlled spatial frequency: see Fig.~\ref{saturation_encoders}~(top), reproduced from~\cite{Li22}. In this case, the slope of the response of these autoencoders (their sensitivity) is bigger for achromatic gratings than for red-green and yellow-blue gratings and it reduces with the contrast of the gratings, just as in humans. 

This contrast-dependent saturation has been described in vision science with specific input-dependent activation functions such as the Divisive Normalization~\cite{Carandini12,Malo10,Laparra10,Schutt17,Martinez18,Martinez19}, which has been found equivalent to classical recurrent models of neural interaction in biology~\cite{Gomez19,Malo24}. Therefore, this behavior, and their adaptation benefits, can be enforced in conventional networks by imposing this kind of interaction in their architecture. Examples include benefits in autoencoding and compression~\cite{Malo06a,Balle17}, denoising and  enhancement~\cite{Gutierrez06,Laparra17}, segmentation~\cite{Hernandez23,Hernandez24},  classification~\cite{Coen13,Bertalmio20,Miller22}, or robustness to adversarial attacks with few layers given the strong nonlinearity due to this biological computation~\cite{Bertalmio20}. 
Moreover, inclusion of these nonlinearities if done parametrically (e.g. by using parametric expressions in the kernel of Divisive Normalization) reduces the training time and increases generalization because of the drastic reduction in the number of parameters of the network~\cite{Vila25b}.

\begin{figure}[b]
\begin{center}
\vspace{-0.3cm}
\includegraphics[width=0.7\textwidth]{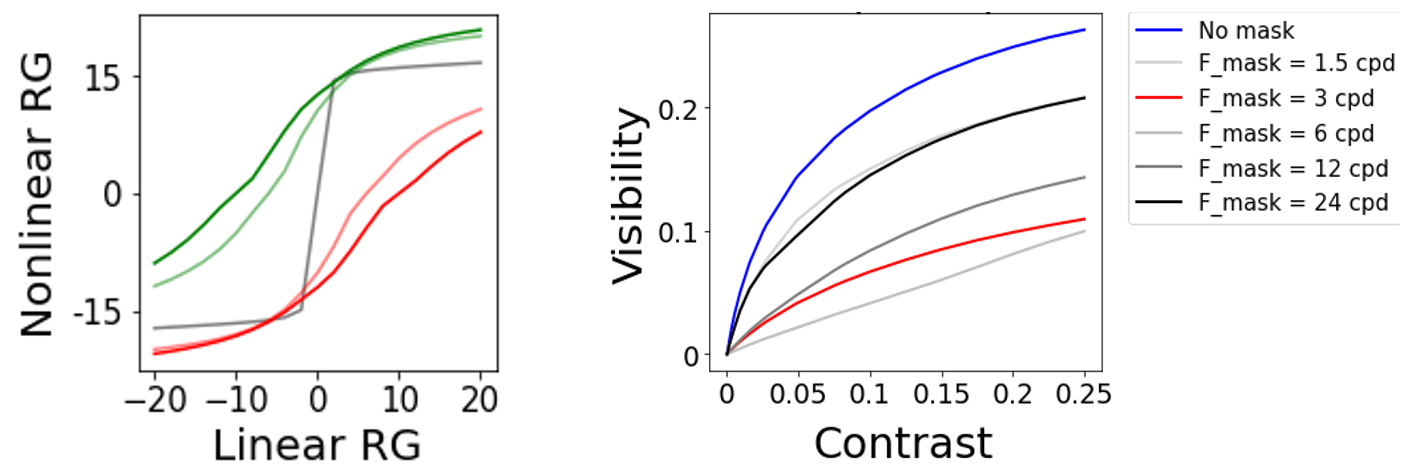}
\vspace{-0.3cm}
\caption{\small{Examples of errors happening in the model~\cite{Martinez18,Martinez19,Malo24}. \emph{Left}: too sharp chroma nonlinearity in achromatic context -wrong gray curve- (taken from Fig.~\ref{color_results}, subplot M). \emph{Right}: too much / too low masking -light gray curves in wrong order- (taken from Fig.~\ref{masking_results}, subplot P). As this particular models is explainable, these could be solved by changing the values of the corresponding kernels of divisive normalization.}}
\label{errors1}
\vspace{-0.9cm}
\end{center}
\end{figure}

As a result, non-human behaviors that are observed in the models using the proposed stimuli and methodology suggest changes in the parameters of the architecture. See for instance the chromatic examples and the texture examples in  Fig.~\ref{errors1} (left and right respectively). The too-sharp behavior of the response in the achromatic illumination condition could be \emph{corrected} through an illumination dependent term in the denominator of the Divisive Normalization. Similarly, in the texture case, the fact that the no-mask condition (blue curve) is below the low-contrast curve suggests that \emph{facilitation} in the Divisive Normalization is too high. These qualitative errors directly suggest quantitative changes in the architecture (if it is formulated in an explainable way), as pointed out in~\cite{Martinez19}.

\subsection{Changes in the optimization goal or training data to get human-like adaptation}

The proposed stimuli and the associated human behaviors may also suggest changes in the optimization goal and on the necessary data to train the networks.

For instance, it is known that information maximization arguments lead to the emergence of Gabor-like receptive fields tuned to achromatic and opponent-chromatic directions~\cite{Hyvarinen09,Gutmann14}.
However, that sensible goal can be complemented with denoising-deblurring tasks so that center-surround cells and proper contrast sensitivity do emerge~\cite{Li92,Karklin11,lindsey19,Li22}.
Moreover, if the contrast nonlinearities do not emerge after all these linear stages, or they are not adaptive enough, this may be enforced by the segmentation goal in the encoder, as in~\cite{Hernandez24}, see Fig.~\ref{saturation_encoders} (bottom).
In that case, the behavior in that segmentation network may not be completely human in part by lack of constraints in the Divisive Normalization (free kernels in~\cite{Hernandez23,Hernandez24} as opposed to more sensible parametric kernels in~\cite{Martinez18,Martinez19}), but part of its adaptive behavior may come from the selection of the training data that enforces contrast adaptation.
Regarding the poor emergence of plausible receptive fields in the considered non-parametric models (see Table~2), this \emph{error} makes sense in the context of the recently proposed \emph{feature-spreading} problem~\cite{Vila25b}: if the goal is not demanding enough (as is usual in conventional goals) the features spread along all layers of the net in a way that the weak goal(s) is (are) fulfilled, but the layers remain biologically non-plausible.  

\begin{figure}[t]
\begin{center}
\vspace{-0.3cm}
\includegraphics[width=1\textwidth]{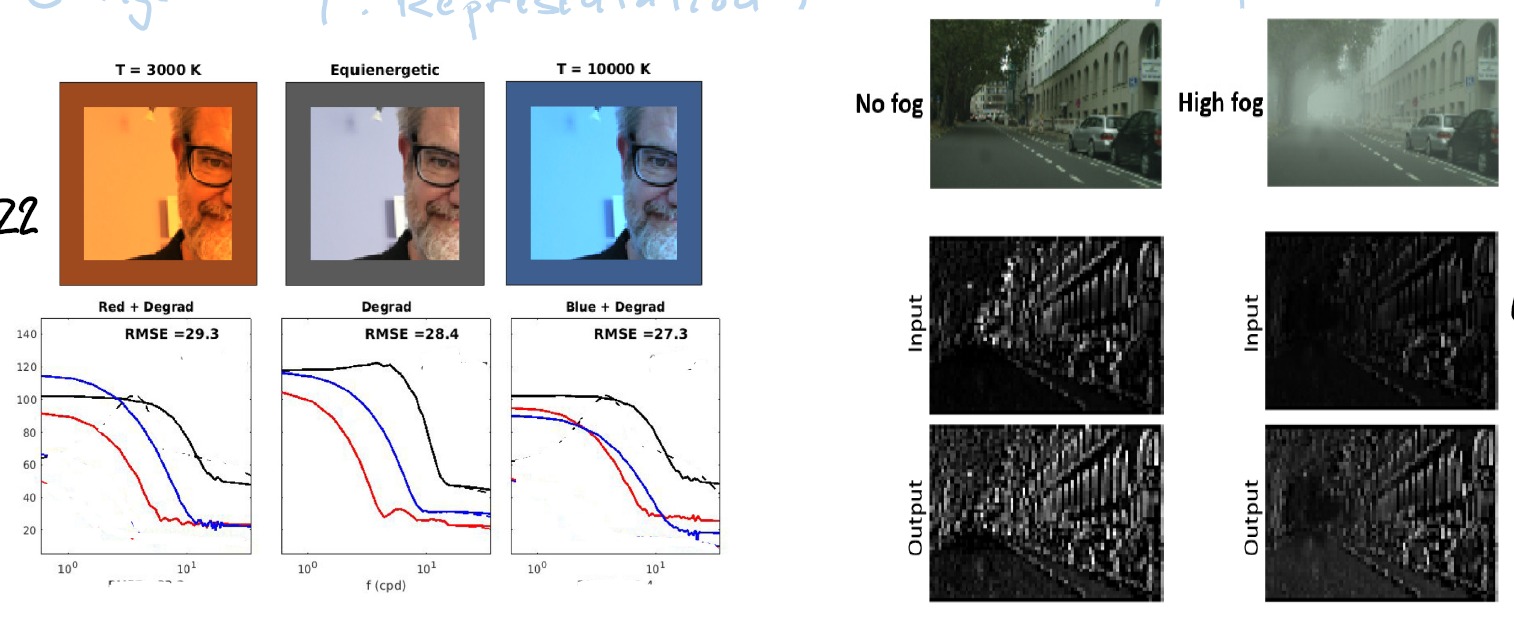}
\vspace{-0.3cm}
\caption{\small{Left: Adaptation in the CSFs in autoencoders obtained from training on the proper (colormetrically calibrated) environments for color adaptation~\cite{Li22}. Right: Improved contrast perception by using bio-inspired Divisive Normalization (with adaptive contrast responses such as the ones described in our proposal) in the model~\cite{Hernandez23}.}}
\label{advantages1}
\vspace{-0.4cm}
\end{center}
\end{figure}

Regarding to suggestions on the training data, the behavior of the achromatic and chromatic CSFs proposed here was checked in~\cite{Li22} under data with well controlled illumination. The behavior found in the autoencoder CSFs in those cases resemble Von Kries adaptation, as was anticipated in~\cite{Gutmann14}. Fig.~\ref{advantages1} (left) shows that under low-temprature (reddish) illumination the red-tuned channel is relatively attenuated with regard to the blue-tuned channel, and the other way around under a high-temperature (blueish) illumination, as would happen using a Von Kries computation~\cite{Fairchild13} or imposing the shifts in the response curves~\cite{Gegen92} shown in the Decalogue.
Finally, Fig.~\ref{advantages1} (right) shows that proper selection of training data (i.e. including images with high fog for segmentation to induce contrast adaptation) leads to the contrast enhancement results, as anticipated by the contrast-dependent nonlinearities shown in Fig.~\ref{saturation_encoders} (bottom). 

\subsection{Human-like curves imply better priors for natural image statistics}

Two examples may illustrate how the nonlinear responses to Gabor stimuli shown in textured contexts as presented in the proposed Decalogue capture the statistics of natural images: the described nonlinear behaviors are a robust prior which may benefit whatever network intended to work in vision.

\begin{figure}[t]
\begin{center}
\vspace{-0.3cm}
\includegraphics[width=0.8\textwidth]{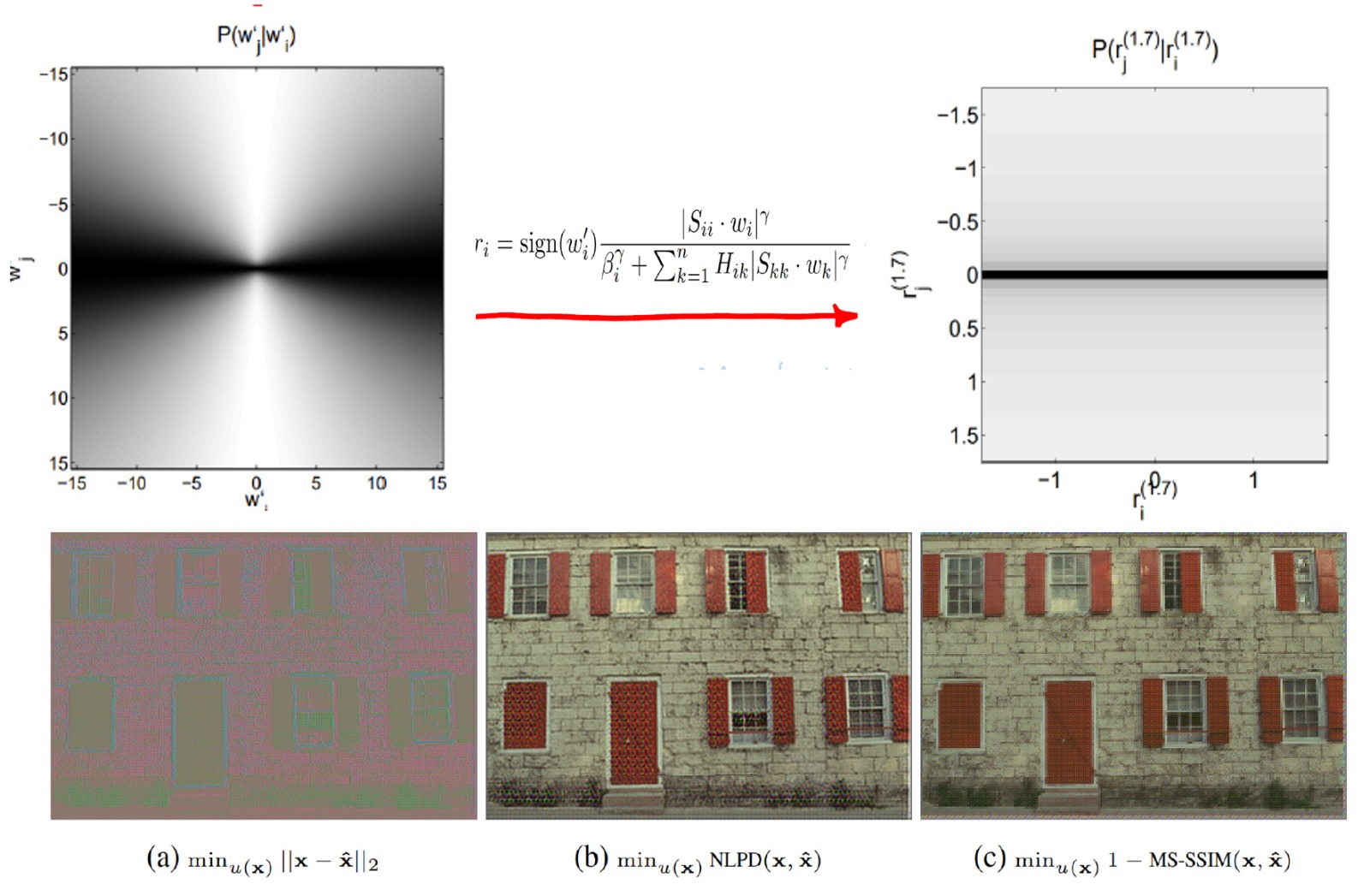}
\vspace{-0.3cm}
\caption{\small{Top: Image PDF factorization from the contrast nonlinearites (div. norm.) illustrated in Figs.~\ref{masking1} and~\ref{masking2}, as shown in~\cite{Malo10}. Bottom: autoencoders trained with distortion metrics based on the contrast nonlinearities described in our proposal capture natural image statistics despite being trained with few samples~\cite{Hepburn22}.}}
\label{advantages2}
\vspace{-0.4cm}
\end{center}
\end{figure}

First, in Fig.~\ref{advantages2} (top) 
we show that the energy of neighbor Gabor-like coefficients is correlated (bow-tie conditional probabilities of Gabor coefficients in natural images), but the nonlinear responses in textured backgrounds make the resulting coefficients independent~\cite{Malo10}. 

Second, non-Euclidean metrics based on the nonlinear responses to the stimuli presented here (e.g. metrics like those reported in~\cite{Laparra10,Laparra17,Martinez19,Malo24}) represent a robust prior of the PDF of natural images as illustrated by the fact shown in Fig.~\ref{advantages2} (bottom): in autoencoders with access to very few samples,
the use of this kind of perceptual metrics,
make the reconstruction of images much more robust than those using (naive) Euclidean metrics because the perceptual metric is already capturing the statistics of natural images although samples are missing~\cite{Hepburn22}.

\section{Final Remarks}

In the first place, we have noted that there are many open problems when we evaluate the human nature of artificial networks: there is a non-trivial relationship between the environment, the task, and the 
architecture~\cite{Poggio21,Malo24a,Hernandez25}. That complexity 
implies it is difficult to choose the layer(s) to measure from and the read-out mechanism to check the human nature of the model responses. These problems point out the need of new tests of human alignment that are independent of the training data and goal.

This motivates our proposal: a set of stimuli, \emph{a Decalogue}, based on classical low-level vision science. The stimuli and associated facts (or human responses) describe the adaptive information bottleneck in the retina-V1 pathway.
Some of the sensitivity surfaces are standardized~\cite{Watson00,Watson02,Mullen85,Daly90,Malo97,Kelly79}, or data is readily available~\cite{Diez11} and allow quantitative comparison (namely properties 1, 3, 4, 6), as done in~\cite{Vila23,Li22}, but the ones that involve showing tests in different illuminations or textured backgrounds (namely properties 2, 7-10) or those related to receptive fields (namely Gabors in 5) have only qualitative value.

The qualitative/quantitative nature of the proposed test is not a problem in practice as shown by its application to analyze and rank three illustrative models: (1) a parametric one based on physiology, classical psychophysics and Maximum Differentiation measurements~\cite{Martinez18,Martinez19,Gomez19,Malo24}, (2) a non-parametric model, the \emph{PerceptNet}~\cite{Hepburn20}, that includes trainable Divisive Normalization to reproduce human opinion on subjective image quality, and 
(3) a U-net with the same encoder as the \emph{PerceptNet} but trained for image segmentation~\cite{Hernandez23,Hernandez24}. The proposed test successfully ranks the models according to their different qualitative origin. It is important to stress that the use of this test can be easily extended to quantitative comparisons (as done in~\cite{Li22,Akbarinia23,Rafal25,Rafal_CVPR_25}), although that is not shown in this presentation.
Even the qualitative application of the test works to rank the human alignment and to point out that the two with less alignment have been trained
for tasks that are not enough to fully explain the human behavior.

Finally, in the discussion, we have seen that the proposed test can be useful to modify the architecture of the networks,
both in their linear and nonlinear parts. The test is useful to question the tasks or restrictions that are used in training (e.g. infoMax, noise, compression bottlenecks, classification, segmentation, etc.). It is also useful to question the data used in the training, either in their generality or balance.
Moreover, we discussed how the use of human behaviors represented by the data in the proposed test, gives rise to priors related to the statistics of natural images.

In summary, we argue that the analysis of any kind of network, not only those that are specifically dedicated to modeling the human vision, but any devoted to vision, can benefit, in great measure, from seeing how they respond to the proposed test.

\enlargethispage{20pt}

\ack{The \emph{invited talk} at the \emph{Artificial Intelligence Evaluation Workshop 2022} was funded be the University of Bristol.
The computational work was partially funded by MCIN/AEI/FEDER/UE under Grants PID2020-118071GB-I00 and PID2023-152133NB-I00, by Spanish MIU under Grant FPU21/02256 and by Generalitat Valenciana under Projects GV/2021/074, CIPROM/2021/056, and by the grant BBVA Foundations of Science program: Maths, Stats, Comp. Sci. and AI (VIS4NN). Some computer resources were provided by Artemisa, funded by the EU ERDF through the Instituto de Física Corpuscular, IFIC (CSIC-UV). The audio-draft of this work (\emph{the talk}) was recorded at El Saler beach (Valencia) and then transcription was done at the \emph{Lisboa} restaurant (Valencia): its staff was particularly helpful at writing time during \emph{Fallas} 2025.}

\bibliographystyle{unsrt}
\bibliography{actual5,biblio_clean}

\begin{thebibliography}{100}

\bibitem{bowers2023deep}
Jeffrey~S Bowers, Gaurav Malhotra, Marin Dujmovi{\'c}, Milton~Llera Montero, Christian Tsvetkov, Valerio Biscione, Guillermo Puebla, Federico Adolfi, John~E Hummel, Rachel~F Heaton, et~al.
\newblock Deep problems with neural network models of human vision.
\newblock {\em Behavioral and Brain Sciences}, 46:e385, 2023.

\bibitem{Bowers24}
V.~Biscione et~al.
\newblock Mind{S}et: Vision. a toolbox for testing {DNN}s on key psychological experiments.
\newblock {\em arXiv preprint arXiv:2404.05290}, 2024.

\bibitem{Rafal_CVPR_25}
Yancheng Cai, Fei Yin, Dounia Hammou, and Rafal Mantiuk.
\newblock Do computer vision foundation models learn the low-level characteristics of the human visual system?
\newblock {\em {CVPR} ArXiV: 2502.20256}, 2025.

\bibitem{Rafal25}
Dounia Hammou, Yancheng Cai, Pavan Madhusudanarao, Christos~G. Bampis, and Rafał~K. Mantiuk.
\newblock Do image and video quality metrics model low-level human vision?
\newblock {\em ArXiV: 2503.16264}, 2025.

\bibitem{ICLR2019}
J.~Kubilius et~al.
\newblock Brain-like object recognition with high-performing shallow recurrent anns.
\newblock {\em ICLR, Arxiv: 1909.06161}, 2019.

\bibitem{PNAS2021}
Johannes Mehrer, Courtney~J. Spoerer, Emer~C. Jones, Nikolaus Kriegeskorte, and Tim~C. Kietzmann.
\newblock An ecologically motivated image dataset for deep learning yields better models of human vision.
\newblock {\em Proc. Nat. Acad. Sci.}, 118(8):e2011417118, 2021.

\bibitem{otroPNAS2021}
C.~Zhuang, S.~Yan, A.~Nayebi, M.~Schrimpf, MC. Frank, JJ. DiCarlo, and DLK. Yamins.
\newblock Unsupervised neural network models of the ventral visual stream.
\newblock {\em Proc. Nat. Acad. Sci.}, 118(3):e2014196118, 2021.

\bibitem{JournalCognSci2021}
K.R. Storrs, T.C. Kietzmann, A.~Walther, J.~Mehrer, and N.~Kriegeskorte.
\newblock Diverse deep neural networks all predict human inferior temporal cortex well, after training and fitting.
\newblock {\em Journal of Cognitive Neuroscience}, 33(10):2044--2064, 2021.

\bibitem{JNeurosci18}
R.~Rajalingham, E.B. Issa, P.~Bashivan, K.~Kar, K.~Schmidt, and J.J. DiCarlo.
\newblock Large-scale, high-resolution comparison of the core visual object recognition behavior of humans, monkeys, and state-of-the-art deep artificial neural networks.
\newblock {\em Journal of Neuroscience}, 38(33):7255--7269, 2018.

\bibitem{NeuralNets2021}
T.~Macpherson, A.~Churchland, T.~Sejnowski, JJ. DiCarlo, Y.~Kamitani, H.~Takahashi, and T.~Hikida.
\newblock Natural and artificial intelligence: A brief introduction to the interplay between ai and neuroscience research.
\newblock {\em Neural Networks}, 144:603--613, 2021.

\bibitem{PLOSCompBiol2019}
SA. Cadena, GH. Denfield, EY. Walker, LA. Gatys, AS. Tolias, M.~Bethge, and S.~Ecker.
\newblock Deep convolutional models improve predictions of macaque v1 responses to natural images.
\newblock {\em PLoS Comput. Biol.}, 15(4):e1006897, 2019.

\bibitem{PLOSCompBiol2020}
MF. Burg, SA. Cadena, Denfield GH., EY. Walker, AS. Tolias, M.~Bethge, and S.~Ecker.
\newblock Learning divisive normalization in primary visual cortex.
\newblock {\em PLoS Comput. Biol.}, 16(6):e1009028, 2021.

\bibitem{PaninskiPersonal01}
L.~Paninsky.
\newblock Personal communication at nyu laboratory for computational vision.
\newblock 2001.

\bibitem{MarrPoggio77}
D.~Marr and T.~Poggio.
\newblock From understanding computation to understanding neural circuitry.
\newblock {\em Neurosci. Res. Prog. Bull.}, 15:470--488, 1977.

\bibitem{Marr78}
D.~Marr.
\newblock {\em Vision: A Computational Investigation into the Human Representation and Processing of Visual Information}.
\newblock W.H. Freeman and Co., New York, 1978.

\bibitem{Poggio21}
Tomaso Poggio.
\newblock From marr{\textquoteright}s vision to the problem of human intelligence.
\newblock {\em {MIT}-{CBMM} Memos}, (118), 09/2021 2021.

\bibitem{Lengyel24}
Máté Lengyel.
\newblock Marr's three levels of analysis are useful as a framework for neuroscience.
\newblock {\em The Journal of Physiology}, 602(9):1911--1914, 2024.

\bibitem{Pillow24}
J.W. Pillow.
\newblock Cross talk opposing view: Marr's three levels of analysis are not useful as a framework for neuroscience.
\newblock {\em The Journal of Physiology}, 602(9):1915--1917, 2024.

\bibitem{Malo24a}
J.~Malo and P.~Hernández-Cámara.
\newblock A separate theory-on-top level may be inspiring, but it is neither separate nor enough.
\newblock {\em The Journal of Physiology.}, 602(9):1918--1918, 2024.

\bibitem{GomezVilla20}
A.~Gomez-Villa, A.~Martin, J.~Vazquez, M.~Bertalm\'io, and J.~Malo.
\newblock Color illusions also deceive {CNNs} for low-level vision tasks: Analysis and implications.
\newblock {\em Vision Research}, 176:156--174, 2020.

\bibitem{Li22}
Qiang Li, Alex Gomez-Villa, Marcelo Bertalmío, and Jesús Malo.
\newblock {Contrast sensitivity functions in autoencoders}.
\newblock {\em Journal of Vision}, 22(6), 2022.

\bibitem{Hernandez25}
Pablo Hernández-Cámara, Jorge Vila-Tomás, Valero Laparra, and Jesús Malo.
\newblock Dissecting the effectiveness of deep features as metric of perceptual image quality.
\newblock {\em Neural Networks}, 185:107189, 2025.

\bibitem{Hepburn22}
Alexander Hepburn, Valero Laparra, Raul Santos-Rodriguez, Johannes Ball{\'e}, and Jesus Malo.
\newblock On the relation between statistical learning and perceptual distances.
\newblock In {\em International Conference on Learning Representations}, 2022.

\bibitem{Malo06b}
J.~Malo and J.~Guti{\'e}rrez.
\newblock V1 non-linear properties emerge from local-to-global non-linear {ICA}.
\newblock {\em Network: Computation in Neural Systems}, 17(1):85--102, 2006.

\bibitem{Laparra12}
V.~Laparra, S.~Jiménez, G.~Camps-Valls, and Jesús Malo.
\newblock Nonlinearities and adaptation of color vision from {S}equential {P}rincipal {C}urves {A}nalysis.
\newblock {\em Neural Comp.}, 24(10):2751--2788, 2012.

\bibitem{Laparra15}
V.~Laparra and J.~Malo.
\newblock Visual aftereffects and sensory nonlinearities from a single statistical framework.
\newblock {\em Frontiers in Human Neuroscience}, 9:557, 2015.

\bibitem{Barlow59}
H.B. Barlow.
\newblock Sensory mechanisms, the reduction of redundancy, and intelligence.
\newblock {\em Proc. of the Nat. Phys. Lab. Symposium on the Mechanization of Thought Process}, (10):535--539, 1959.

\bibitem{Barlow61}
H.B. Barlow.
\newblock Possible principles underlying the transformation of sensory messages.
\newblock In WA~Rosenblith, editor, {\em Sensory Communication}, pages 217--234. MIT Press, Cambridge, MA, 1961.

\bibitem{Barlow01}
H.B. Barlow.
\newblock Redundancy reduction revisited.
\newblock {\em {Network:} {Computation} in Neural Systems}, 12:241--253, 2001.

\bibitem{GRCmeetingOxford04}
J.~Malo, J.~Gutiérrez, and J.~Rovira.
\newblock Perturbation analysis of the changes in {V1} receptive fields due to context.
\newblock {\em Gordon Research Conference: Sensory Coding and the Natural Environment}, 2004.

\bibitem{BarlowPersonal04}
H.~Barlow.
\newblock Personal communication at the {GRC} sens. coding nat. env.
\newblock 2004.

\bibitem{Malo15}
Jes{\'u}s {Malo} and Eero~P. {Simoncelli}.
\newblock Geometrical and statistical properties of vision models obtained via maximum differentiation.
\newblock In Bernice~E. {Rogowitz}, Thrasyvoulos~N. {Pappas}, and Huib {de Ridder}, editors, {\em Human Vision and Electronic Imaging XX}, volume 9394 of {\em Society of Photo-Optical Instrumentation Engineers (SPIE) Conference Series}, page 93940L, March 2015.

\bibitem{Martinez18}
M.~Martinez, P.~Cyriac, T.~Batard, M.~Bertalm{\'\i}o, and J.~Malo.
\newblock Derivatives and inverse of cascaded linear+nonlinear neural models.
\newblock {\em PLOS ONE}, 13(10):1--49, 10 2018.

\bibitem{Martinez19}
M.~Martinez, M~Bertalm\'io, and J.~Malo.
\newblock In praise of artifice reloaded: Caution with natural image databases in modeling vision.
\newblock {\em Front. Neurosci. doi: 10.3389/fnins.2019.00008}, 2019.

\bibitem{Malo24}
J.~Malo, JJ. Esteve-Taboada, and M~Bertalmío.
\newblock Cortical divisive normalization from wilson-cowan neural dynamics.
\newblock {\em J. Nonlinear Sci.}, 34(2):35, 2024.

\bibitem{Hepburn20}
A.~{Hepburn}, V.~{Laparra}, J.~{Malo}, R.~{McConville}, and R.~{Santos-Rodriguez}.
\newblock Perceptnet: A human visual system inspired neural network for estimating perceptual distance.
\newblock In {\em IEEE ICIP}, pages 121--125, 2020.

\bibitem{Balle17}
J.~Ballé, V.~Laparra, and EP. Simoncelli.
\newblock End-to-end optimized image compression.
\newblock {\em ICLR ArXiV:1611.01704}, 2017.

\bibitem{Hernandez23}
P.~Hernández-Cámara, J.~Vila-Tomás, V.~Laparra, and J.~Malo.
\newblock Neural networks with divisive normalization for image segmentation.
\newblock {\em Patt. Recogn. Lett.}, 173:64--71, 2023.

\bibitem{Hernandez24}
P.~Hernández-Cámara, J.~Vila-Tomás, P.~Dauden-Oliver, N.~Alabau-Bosque, V.~Laparra, and J.~Malo.
\newblock Why divisive normalization works in image segmentation?
\newblock {\em Neurocomputing}, 649, 2025.

\bibitem{Torralba24}
Antonio Torralba, Phillip Isola, and William~T Freeman.
\newblock {\em Foundations of Computer Vision}.
\newblock MIT Press, 2024.

\bibitem{Li92}
J.~J. {Atick}, Z.~{Li}, and A.~N. {Redlich}.
\newblock Understanding retinal color coding from first principles.
\newblock {\em Neural Computation}, 4(4):559--572, 1992.

\bibitem{Karklin11}
Y.~Karklin and E.~Simoncelli.
\newblock Efficient coding of natural images with a population of noisy linear-nonlinear neurons.
\newblock In {\em Advances in Neural Information Processing Systems}, volume~24. Curran Associates, Inc., 2011.

\bibitem{lindsey19}
J.~Lindsey, SA. Ocko, S.~Ganguli, and S.~Deny.
\newblock The effects of neural resource constraints on early visual representations.
\newblock {\em Int. Conf. Learn. Repr. ICLR}, 2019.

\bibitem{Li14}
Li~Zhaoping.
\newblock {\em Understanding Vision: Theory, Models, and Data}.
\newblock Oxford University Press, 05 2014.

\bibitem{Wallace92}
Gregory~K. Wallace.
\newblock The {JPEG} still picture compression standard.
\newblock {\em Commun. ACM}, 34(4):30–44, April 1991.

\bibitem{Malo95}
J.~Malo, A.M. Pons, and J.M. Artigas.
\newblock Bit allocation algorithm for codebook design in vector quantization fully based on human visual system non-linearities for suprathreshold contrasts.
\newblock {\em Electronics Letters}, 31(15):1229--1231, 1995.

\bibitem{Malo00b}
J.~Malo, F.~Ferri, J.~Albert, J.Soret, and J.M. Artigas.
\newblock The role of perceptual contrast non-linearities in image transform coding.
\newblock {\em Image \& Vision Computing}, 18(3):233--246, 2000.

\bibitem{Marcellin01}
David Taubman and Michael Marcellin.
\newblock {\em JPEG2000 Image Compression Fundamentals, Standards and Practice}.
\newblock Springer Publishing Company, Incorporated, 2013.

\bibitem{Malo06a}
J.~Malo, I.~Epifanio, R.~Navarro, and E.~Simoncelli.
\newblock Non-linear image representation for efficient perceptual coding.
\newblock {\em IEEE Transactions on Image Processing}, 15(1):68--80, 2006.

\bibitem{Legall92}
Didier~J. {Le Gall}.
\newblock The {MPEG} video compression algorithm.
\newblock {\em Signal Processing: Image Communication}, 4(2):129--140, 1992.

\bibitem{Malo00a}
J.~Malo, F.~Ferri, J.~Gutierrez, and I.~Epifanio.
\newblock Importance of quantizer design compared to optimal multigrid motion estimation in video coding.
\newblock {\em Electronics Letters}, 36(9):507--509, 2000.

\bibitem{Malo00c}
J.~Malo, J.~Gutierrez, I.~Epifanio, and F.~Ferri.
\newblock Perceptually weighted optical flow for motion-based segmentation in {MPEG}-4 paradigm.
\newblock {\em Electronics Letters}, 36(20):1693--1694, 2000.

\bibitem{Malo01a}
J.Malo, J.Gutierrez, I.Epifanio, F.Ferri, and J.M.Artigas.
\newblock Perceptual feed-back in multigrid motion estimation using an improved {DCT} quantization.
\newblock {\em IEEE Transactions on Image Processing}, 10(10):1411--1427, 2001.

\bibitem{Goodale91}
MA. Goodale, AD.; Milner, LS. Jakobson, and DP. Carey.
\newblock A neurological dissociation between perceiving objects and grasping them.
\newblock {\em Nature}, 349(6305):154--156, 1991.

\bibitem{Milner92}
A.~D. Milner and M.~A. Goodale.
\newblock Separate visual pathways for perception and action.
\newblock {\em Trends Neurosci.}, 15:20--25, 1992.

\bibitem{Logothetis96}
NK. Logothetis and DL. Sheinberg.
\newblock Visual object recognition.
\newblock {\em Ann. Rev. Neurosci.}, 19:577--621, 1996.

\bibitem{Koch00}
G.~Kreiman, C.~Koch, and I.~Fried.
\newblock Category specific visual responses of single neurons in the human medial temporal lobe.
\newblock {\em Nat. Neurosci.}, 3(9):946--953, 2000.

\bibitem{Coen13}
Ruben Coen-Cagli and Odelia Schwartz.
\newblock The impact on midlevel vision of statistically optimal divisive normalization in v1.
\newblock {\em Journal of Vision}, 13(8):13--13, 07 2013.

\bibitem{Miller22}
Michelle Miller, SueYeon Chung, and Kenneth~D. Miller.
\newblock Divisive feature normalization improves image recognition performance in alexnet.
\newblock In {\em Int. Conf. Learn. Repres. ICLR}, 2022.

\bibitem{Akbarinia23}
A.~Akbarinia, Y.~Morgenstern, and K.R. Gegenfurtner.
\newblock Contrast sensitivity function in deep networks.
\newblock {\em Neural Networks}, 164:228--244, 2023.

\bibitem{CCN25}
P~Hernández-Cámara, A~Gomez-Villa, JoseManuel Jaén-Lorites, J~Vila-Tomás, J~Malo, and V~Laparra.
\newblock Contrast sensitivity function of multimodal vision-language models.
\newblock In {\em 8th Cognitive Computational Neuroscience Conference}, 2025.

\bibitem{Teo94a}
P.C. Teo and D.J. Heeger.
\newblock Perceptual image distortion.
\newblock {\em Proceedings of the SPIE}, 2179:127--141, 1994.

\bibitem{Duda73}
R.O. Duda and P.E. Hart.
\newblock {\em Pattern Classification and Scene Analysis}.
\newblock John Wiley \& Sons, New York, 1973.

\bibitem{graham_visual_1989}
Norma V.~S. Graham.
\newblock {\em Visual pattern analyzers}.
\newblock Visual pattern analyzers. Oxford University Press, New York, NY, US, 1989.
\newblock Pages: xvi, 646.

\bibitem{Laparra10}
V.~Laparra, J.~Mu\~{n}oz Mar\'i, and J.~Malo.
\newblock Divisive normalization image quality metric revisited.
\newblock {\em JOSA A}, 27(4):852--864, 2010.

\bibitem{Wang03}
Z~Wang, A~C Bovik, H~R Sheikh, and E~P Simoncelli.
\newblock Perceptual image quality assessment: {From} error visibility to structural similarity.
\newblock {\em IEEE Trans Image Processing}, 13(4):600--612, 2004.

\bibitem{Ding20}
Keyan Ding, Kede Ma, Shiqi Wang, and Eero~P. Simoncelli.
\newblock Image quality assessment: Unifying structure and texture similarity.
\newblock {\em IEEE Transactions on Pattern Analysis and Machine Intelligence}, 44(5):2567--2581, 2022.

\bibitem{Sheikh05}
H.R. Sheikh, A.C. Bovik, and G.~de~Veciana.
\newblock An information fidelity criterion for image quality assessment using natural scene statistics.
\newblock {\em IEEE Transactions on Image Processing}, 14(12):2117--2128, 2005.

\bibitem{Sheikh06}
H.R. Sheikh and A.C. Bovik.
\newblock Image information and visual quality.
\newblock {\em IEEE Transactions on Image Processing}, 15(2):430--444, 2006.

\bibitem{Malo20}
J.~Malo.
\newblock Spatio-chromatic information available from different neural layers via gaussianization.
\newblock {\em The Journal of Mathematical Neuroscience}, 10(18), 2020.

\bibitem{Malo21}
J.~Malo, B.~Kheravdar, and Q.~Li.
\newblock Visual information fidelity with better vision models and better mutual information estimates.
\newblock {\em Journal of Vision}, 21(9):2351, 2021.

\bibitem{Li24}
Qiang Li, Greg~Ver Steeg, and Jesus Malo.
\newblock Functional connectivity via total correlation: Analytical results in visual areas.
\newblock {\em Neurocomputing}, 571:127143, 2024.

\bibitem{Gatys16}
Leon~A. Gatys, Alexander~S. Ecker, and Matthias Bethge.
\newblock Image style transfer using convolutional neural networks.
\newblock In {\em 2016 IEEE Conference on Computer Vision and Pattern Recognition (CVPR)}, pages 2414--2423, 2016.

\bibitem{Laparra24}
V.~Laparra, JE. Johnson, G.~Camps-Valls, R.~Santos-Rodriguez, and J.~Malo.
\newblock { Estimating Information Theoretic Measures via Multidimensional Gaussianization}.
\newblock {\em IEEE Trans. Patt. Anal. \& Mach. Intell.}, 47(02):1293--1308, 2025.

\bibitem{Malo25}
J.~Malo, JJ. Esteve-Taboada, G.~Aguilar, M.~Maertens, and FA. Wichmann.
\newblock Estimating the contribution of early and late noise in vision from psychophysical data.
\newblock {\em J. Vision}, 25(1):12--12, 2025.

\bibitem{Vedaldi16}
A.~Mahendran and A.~Vedaldi.
\newblock Visualizing deep convolutional neural networks using natural pre-images.
\newblock {\em Int. J. Comput. Vis.}, 120:233–255, 2016.

\bibitem{Luo16}
W.~Luo, Y.~Li, R.~Urtasun, and R.~Zemel.
\newblock Understanding the effective receptive field in deep convolutional neural networks.
\newblock In {\em Advances in Neural Information Processing Systems}, volume~29. Curran Associates, Inc., 2016.

\bibitem{Hubel59}
David~H Hubel, Torsten~N Wiesel, et~al.
\newblock Receptive fields of single neurones in the cat’s striate cortex.
\newblock {\em J physiol}, 148(3):574--591, 1959.

\bibitem{Hubel61}
David~H Hubel and Torsten~N Wiesel.
\newblock Integrative action in the cat's lateral geniculate body.
\newblock {\em The Journal of physiology}, 155(2):385, 1961.

\bibitem{Ringach02}
D.L. Ringach.
\newblock Spatial structure and symmetry of simple-cell receptive fields in macaque primary visual cortex.
\newblock {\em J. Neurophysiol.}, 88(1):455‐463, 2002.

\bibitem{Lennie08}
C.~Tailby, SG. Solomon, NT. Dhruv, and P.~Lennie.
\newblock Habituation reveals fundamental chromatic mechanisms in striate cortex of macaque.
\newblock {\em J. Neurosci.}, 28(5):1131--1139, 2008.

\bibitem{Ringach04}
D.~Ringach and R.~Shapley.
\newblock Reverse correlation in neurophysiology.
\newblock {\em Cognit. Sci.}, 28(2):147--166, 2004.
\newblock Rendering the Use of Visual Information from Spiking Neurons to Recognition.

\bibitem{Eckstein02}
MP. Eckstein and AJ. Ahumada.
\newblock Classification images: A tool to analyze visual strategies.
\newblock {\em J. Vision}, 2(1), 2002.

\bibitem{Stiles00}
G~Wyszecki and WS. Stiles.
\newblock {\em Color Science: Concepts and Methods, Quantitative Data and Formulae}.
\newblock John Wiley \& Sons, New Jersey, 2000.

\bibitem{Otazu10}
X.~Otazu, CA. Parraga, and M.~Vanrell.
\newblock Toward a unified chromatic induction model.
\newblock {\em J. Vision}, 10(12):5--5, 10 2010.

\bibitem{WareCowan82}
C.~Ware and WB. Cowan.
\newblock Changes in perceived color due to chromatic interactions.
\newblock {\em Vision Research}, 22(11):1353--1362, 1982.

\bibitem{gomez-villa_color_2020}
A.~Gomez-Villa, A.~Martín, J.~Vazquez-Corral, M.~Bertalmío, and J.~Malo.
\newblock Color illusions also deceive {CNNs} for low-level vision tasks: {Analysis} and implications.
\newblock {\em Vision Research}, 176:156--174, November 2020.

\bibitem{GomezVilla25}
A.~Gomez-Villa, K.~Wang, CA. Parraga, B.~Twardowski, J.~Malo, J.~Vazquez-Corral, and J.~van~de Weijer.
\newblock The art of deception: Color visual illusions and diffusion models.
\newblock {\em IEEE Comp. Vis. Patt. Recogn. (CVPR)}, 2025.

\bibitem{Malo24charla}
J.~Malo and J.~Bowers.
\newblock The low-level mindset: compelling low-level visual psychophysics to evaluate image computable vision models.
\newblock Invited talk, Psychol. Dept. University of Bristol, 2024.

\bibitem{Rust06}
NC. Rust and JA. Movshon.
\newblock In praise of artifice.
\newblock {\em Nature Neurosci.}, 8(12):1647--1650, 2005.

\bibitem{Schutt17}
HH. Schütt and FA. Wichmann.
\newblock An image-computable psychophysical spatial vision model.
\newblock {\em J. Vision}, 17(12):12--12, 10 2017.

\bibitem{Bertalmio20}
M.~Bertalmío, A.~Gomez-Villa, A.~Martín, J.~Vazquez, D.~Kane, and J.~Malo.
\newblock Evidence for the intrinsically nonlinear nature of receptive fields in vision.
\newblock {\em Scientific Reports}, 10:16277, 2020.

\bibitem{Bertalmio24}
M.~Bertalmío, A.~Durán-Vizcaíno, J.~Malo, and FA. Wichmann.
\newblock Plaid masking explained with input-dependent dendritic nonlinearities.
\newblock {\em Sci. Rep.}, 14:24856, 2024.

\bibitem{Modelfest}
T.~Carney, SA. Klein, CW. Tyler, AD. Silverstein, B.~Beutter, D.~Levi, AB. Watson, AJ. Reeves, AM. Norcia, C.~Chen, W.~Makous, and MP. Eckstein.
\newblock {Development of an image/threshold database for designing and testing human vision models}.
\newblock In {\em Human Vision and Electronic Imaging IV}, volume 3644, pages 542 -- 551. International Society for Optics and Photonics, SPIE, 1999.

\bibitem{BrainScore}
Martin Schrimpf, Jonas Kubilius, Ha~Hong, Najib~J. Majaj, Rishi Rajalingham, Elias~B. Issa, Kohitij Kar, Pouya Bashivan, Jonathan Prescott-Roy, Franziska Geiger, Kailyn Schmidt, Daniel L.~K. Yamins, and James~J. DiCarlo.
\newblock Brain-score: Which artificial neural network for object recognition is most brain-like?
\newblock {\em bioRxiv preprint}, 2018.

\bibitem{Daly93}
S.~Daly.
\newblock Visible differences predictor: An algorithm for the assessment of image fidelity.
\newblock In A.B. Watson, editor, {\em Digital Images and Human Vision}, pages 179--206, Massachusetts, 1993. MIT Press.

\bibitem{Watson93libro}
A.B. Watson et~al.
\newblock {\em Digital Images and Human Vision}.
\newblock MIT Press, Massachusetts, 1993.

\bibitem{Malo97}
J.~Malo, A.M. Pons, and J.M. Artigas.
\newblock Subjective image fidelity metric based on bit allocation of the human visual system in the dct domain.
\newblock {\em Image and Vision Computing}, 15(7):535--548, 1997.

\bibitem{Watson02}
A.~B. {Watson} and J.~{Malo}.
\newblock Video quality measures based on the standard spatial observer.
\newblock In {\em {IEEE} Proc. Int. Conf. Im. Proc.}, volume~3, pages III--III, 2002.

\bibitem{Hernandez24front}
P.~Hernández-Cámara, P.~Daudén-Oliver, V.~Laparra, and J.~Malo.
\newblock Alignment of color discrimination in humans and image segmentation networks.
\newblock {\em Front. Psychol.}, Volume 15 - 2024, 2024.

\bibitem{Olshausen96}
B.~Olshausen and D.~Field.
\newblock Emergence of simple-cell receptive field properties by learning a sparse code for natural images.
\newblock {\em Nature}, 281:607--609, 1996.

\bibitem{Schwartz01}
O.~Schwartz and E.P. Simoncelli.
\newblock Natural signal statistics and sensory gain control.
\newblock {\em Nat. Neurosci.}, 4(8):819--825, 2001.

\bibitem{Malo10}
J.~Malo and V.~Laparra.
\newblock Psychophysically tuned divisive normalization approximately factorizes the pdf of natural images.
\newblock {\em Neural computation}, 22(12):3179--3206, 2010.

\bibitem{Gomez19}
Alexander Gomez-Villa, Marcelo Bertalm\'{\i}o, and Jesus Malo.
\newblock Visual information flow in {W}ilson–{C}owan networks.
\newblock {\em Journal of Neurophysiology}, 123(6):2249--2268, 2020.

\bibitem{Malo22}
J.~Malo.
\newblock Information flow in biological networks for color vision.
\newblock {\em Entropy}, 24:1442, 2022.

\bibitem{Jameson57}
Leo~M. Hurvich and Dorothea Jameson.
\newblock An opponent-process theory of color vision.
\newblock {\em Psychological Review}, 64, Part 1 6:384--404, 1957.

\bibitem{Campbell68}
F.W. Campbell and J.G. Robson.
\newblock Application of {F}ourier analysis to the visibility of gratings.
\newblock {\em Journal of Physiology}, 197:551--566, 1968.

\bibitem{Mullen85}
K.~T. Mullen.
\newblock The {CSF} of human colour vision to red-green and yellow-blue chromatic gratings.
\newblock {\em J. Physiol.}, 359:381--400, 1985.

\bibitem{Georgeson75}
MA. Georgeson and GD. Sullivan.
\newblock Contrast constancy: deblurring in human vision by spatial frequency channels.
\newblock {\em J. Physiol.}, 252(3):627--656, 1975.

\bibitem{Daly90}
S.~Daly.
\newblock Application of a noise-adaptive {Contrast} {Sensitivity} {Function} to image data compression.
\newblock {\em Optical Engineering}, 29(8):977--987, 1990.

\bibitem{Kelly79}
D.~H. Kelly.
\newblock Motion and vision. ii. stabilized spatio-temporal threshold surface.
\newblock {\em J. Opt. Soc. Am.}, 69(10):1340--1349, Oct 1979.

\bibitem{Diez11}
MA. Díez-Ajenjo, P.~Capilla, and MJ. Luque.
\newblock Red-green vs. blue-yellow spatio-temporal contrast sensitivity across the visual field.
\newblock {\em J. Mod. Opt.}, 58(19-20):1736--1748, 2011.

\bibitem{Fairchild13}
M.D. Fairchild.
\newblock {\em Color Appearance Models}.
\newblock The Wiley-IS\&T Series in Imaging Science and Technology. Wiley, 2013.

\bibitem{Whittle92}
Paul Whittle.
\newblock Brightness, discriminability and the “crispening effect”.
\newblock {\em Vision Research}, 32(8):1493--1507, 1992.

\bibitem{Laughlin81}
S.~Laughlin.
\newblock A simple coding procedure enhances a neuron’s information capacity.
\newblock {\em Zeitschrift Für Naturforschung C}, 36(9):910--912, 1981.

\bibitem{Colorlab02}
J.~Malo and M.J. Luque.
\newblock {ColorLab: A Matlab Toolbox for Color Science and Calibrated Color Image Processing}.
\newblock {\em Univ. Valencia. https://isp.uv.es/code/vision\_and\_color/colorlab/content/}, 2002.

\bibitem{Vila23}
J.~Vila-Tomás, P.~Hernández-Cámara, and J.~Malo.
\newblock Artificial psychophysics questions classical hue cancellation experiments.
\newblock {\em Frontiers in Neuroscience}, 17, 2023.

\bibitem{Gegen92}
J.~Krauskopf and K.~Gegenfurtner.
\newblock Color discrimination and adaptation.
\newblock {\em Vision Research}, 32(11):2165--2175, 1992.

\bibitem{Hita93}
Javier Romero, Jos\'{e}~A. Garc\'{i}a, Luis~Jim\'{e}nez del Barco, and E.~Hita.
\newblock Evaluation of color-discrimination ellipsoids in two-color spaces.
\newblock {\em J. Opt. Soc. Am. A}, 10(5):827--837, May 1993.

\bibitem{Legge80}
G.E Legge and J.M. Foley.
\newblock Contrast masking in human vision.
\newblock {\em Journal of the Optical Society of America}, 70:1458--1471, 1980.

\bibitem{Legge81}
G.E Legge.
\newblock A power law for contrast discrimination.
\newblock {\em Vision Research}, 18:68--91, 1981.

\bibitem{Foley94}
John~M. Foley.
\newblock Human luminance pattern-vision mechanisms: masking experiments require a new model.
\newblock {\em J. Opt. Soc. Am. A}, 11(6):1710--1719, Jun 1994.

\bibitem{Watson97}
Andrew~B. Watson and Joshua~A. Solomon.
\newblock Model of visual contrast gain control and pattern masking.
\newblock {\em J. Opt. Soc. Am. A}, 14(9):2379--2391, Sep 1997.

\bibitem{Vistalab}
J.~Malo and J.~Gutierrez.
\newblock {VistaLab: The Matlab toolbox for linear spatio-temporal Vision Models}.
\newblock {\em Univ. Valencia. https://isp.uv.es/code/vision\_and\_color/colorlab/vistalab/}, 2002.

\bibitem{Ross91}
John Ross, Harriet~D. Speed, and Fergus~William Campbell.
\newblock Contrast adaptation and contrast masking in human vision.
\newblock {\em Proceedings of the Royal Society of London. Series B: Biological Sciences}, 246(1315):61--70, 1991.

\bibitem{Shapley11}
Robert Shapley and Michael~J. Hawken.
\newblock Color in the cortex: single- and double-opponent cells.
\newblock {\em Vision Research}, 51(7):701--717, 2011.
\newblock Vision Research 50th Anniversary Issue: Part 1.

\bibitem{Blakemore69}
C.~Blakemore and F.~Campbell.
\newblock On the existence of neurons selectivity sensitive to the orientation and size of retinal images.
\newblock {\em J. Physiol.}, 203:237–260, 1969.

\bibitem{Martinez17}
M.~Martinez, LM. Martinez, and J.~Malo.
\newblock Topographic independent component analysis reveals random scrambling of orientation in visual space.
\newblock {\em {PLoS} {ONE}}, 12(6):e0178345, 2017.

\bibitem{Loxley17}
P.~N. Loxley.
\newblock The two-dimensional gabor function adapted to natural image statistics: A model of simple-cell receptive fields and sparse structure in images.
\newblock {\em Neural Computation}, 29(10):2769--2799, 2017.

\bibitem{Gutmann14}
M.~U. Gutmann, V.~Laparra, A.~Hyv{\"a}rinen, and J.~Malo.
\newblock Spatio-chromatic adaptation via higher-order canonical correlation analysis of natural images.
\newblock {\em {PloS} {ONE}}, 9(2):e86481, 2014.

\bibitem{Carandini94}
M.~Carandini and D.~Heeger.
\newblock Summation and division by neurons in visual cortex.
\newblock {\em Science}, 264(5163):1333--6, 1994.

\bibitem{Carandini12}
Matteo Carandini and David~J. Heeger.
\newblock Normalization as a canonical neural computation.
\newblock {\em Nature Reviews Neuroscience}, 13(1):51--62, January 2012.
\newblock Number: 1 Publisher: Nature Publishing Group.

\bibitem{Alexnet}
Alex Krizhevsky, Ilya Sutskever, and Geoffrey~E Hinton.
\newblock Imagenet classification with deep convolutional neural networks.
\newblock In F.~Pereira, C.J. Burges, L.~Bottou, and K.Q. Weinberger, editors, {\em Advances in Neural Information Processing Systems}, volume~25. Curran Associates, Inc., 2012.

\bibitem{Laparra17}
Valero Laparra, Alexander Berardino, Johannes Ball\'{e}, and Eero~P. Simoncelli.
\newblock Perceptually optimized image rendering.
\newblock {\em J. Opt. Soc. Am. A}, 34(9):1511--1525, Sep 2017.

\bibitem{Vila25b}
Jorge Vila-Tomás, Pablo Hernández-Cámara, Valero Laparra, and Jesús Malo.
\newblock Parametric perceptnet: A bio-inspired deep-net trained for image quality assessment.
\newblock {\em ArXiV}, page 2412.03210, 2025.

\bibitem{Simoncelli90}
E.P. Simoncelli and E.H. Adelson.
\newblock {\em Subband Image Coding}, chapter Subband Transforms, pages 143--192.
\newblock Kluwer Academic Publishers, Norwell, MA, 1990.

\bibitem{DeAngelis97}
D.~Cai, GC. DeAngelis, and RD. Freeman.
\newblock Spatiotemporal receptive field organization in the lateral geniculate nucleus of cats and kittens.
\newblock {\em J. Neurophysiol.}, 78(2):1045--1061, 1997.

\bibitem{Watson83}
A.B. Watson.
\newblock Detection and recognition of simple spatial forms.
\newblock In O.J. Braddick and A.C. Sleigh, editors, {\em Physical and Biological Processing of Images}, volume~11 of {\em Springer Series on Information Sciences}, pages 100--114, Berlin, 1983. Springer Verlag.

\bibitem{Uriegas97}
E.~Martinez-Uriegas.
\newblock Color detection and color contrast discrimination thresholds.
\newblock In {\em Proc. OSA Meeting}, page~81, 1997.

\bibitem{Gutierrez06}
J.~Guti\'errez, F.~Ferri, and J.~Malo.
\newblock Regularization operators for natural images based on nonlinear perception models.
\newblock {\em IEEE Tr. Im. Proc.}, 15(1):189--200, 2006.

\bibitem{Hyvarinen09}
A.~Hyvarinen, J.~Hurri, and PO. Hoyer.
\newblock {\em Natural Image Statistics: A Probabilistic Approach to Early Computational Vision}.
\newblock Springer, 2009.

\bibitem{Watson00}
A.B. Watson and C.V. Ramirez.
\newblock A {S}tandard {O}bserver for {S}patial {V}ision.
\newblock {\em Investig. Opht. and Vis. Sci.}, 41(4):S713, 2000.

\end{thebibliography}

\end{document}